\documentclass{article}
\usepackage{arxiv} 
\usepackage[T1]{fontenc}   
\usepackage[utf8]{inputenc}
\usepackage[english]{babel}
\usepackage[numbers]{natbib}   
\setcitestyle{authoryear,open={(},close={)}}
\usepackage{graphicx} 
\usepackage{enumerate}
\usepackage{multirow}
\usepackage{hyperref}       
\usepackage{url}            
\usepackage{hyperref}
\hypersetup{
	colorlinks=true,
	linkcolor=blue,
	filecolor=blue,
	citecolor = blue,      
	urlcolor=cyan,
}
\usepackage{booktabs}       
\usepackage{multicol}
\usepackage{amsmath,amsfonts,amssymb}
\usepackage{amsthm}        
\numberwithin{equation}{subsection}
\usepackage{nicefrac}      
\usepackage{microtype}     
\usepackage{lipsum}
\usepackage{float} 
\usepackage{bm}
\usepackage{adjustbox}
\usepackage[linesnumbered,lined,boxed,commentsnumbered]{algorithm2e}

\newcommand{\indic}{\mathrm{\textit{I}}}

\newcommand{\Esp}[1]{\mathrm{E}\! \left[ #1 \right]}

\newcommand{\indep}{\perp \!\!\! \perp}
\newcommand{\nindep}{\not\!\perp\!\!\!\perp }

\usepackage{caption}
\usepackage{tikz}
\usetikzlibrary{shapes,arrows}
\usetikzlibrary{decorations.pathreplacing}

\title{Bonus-Malus Scale Premiums for Tweedie's Compound Poisson Models}

\author{
  Jean-Philippe Boucher \& Raïssa Coulibaly \\
  Chaire Co-operators en analyse des risques actuariels\\
  Departement de mathematiques\\
  Universite du Quebec a Montreal \\
}

\begin{document}

\newpage
\maketitle

\begin{abstract}
Based on the recent paper by \cite{delong2020}, two distributions for the total claims amount (loss cost) are considered: Compound Poisson-gamma (CPG) and Tweedie. Each is used as an underlying distribution in the Bonus-Malus Scale (BMS) model described in the paper by \citet{boucher2022}. The BMS model links the premium of an insurance contract to a function of the insurance experience of the related policy. In other words, the idea is to model the increase and the decrease in premiums for insureds that do or do not file claims. Therefore, our proposed models can be seen as a generalization of the paper  of \citet{delong2020} and an extension of the work of \citet{boucher2022}. We applied our approach to a sample of data from a major insurance company in Canada. Data fit and predictability were analyzed. We showed that the studied models are exciting alternatives to consider from a practical point of view, and that predictive ratemaking models can address some important practical considerations.
\end{abstract}

\keywords{Panel data, experience rating, bonus-malus scale, compound poisson-gamma, tweedie, elastic-net}

\section{Introduction}

Experience Rating and predictive ratemaking refer to ratemaking models that use claims information from past insurance contracts to predict the future total amount of claims (also known as "loss costs"). From a ratemaking point-of-view, the idea of experience rating is to compute a premium for insured $i$, for a contract of period $T$, that will consider all the insured’s past insurance contracts $1, \ldots, T-1$. 

Historically, research on this type of predictive ratemaking has focused on modeling the annual claims number of a contract. 
For an insured $i$, one can use panel data theory to model the joint distribution of the annual claims number for each $T$ contract . This is expressed as the product of predictive distributions:

\begin{eqnarray*}
\Pr(N_{i,1} = n_{i,1}, \ldots, N_{i,T}= n_{i,T}) = \prod_{t=1}^T \Pr(N_{i,t}| \boldsymbol{n}_{i,(1:t-1)}) \ ,
\label{produni}
\end{eqnarray*}

where $\boldsymbol{n}_{i,(1:t-1)} = \{n_{i,1}, n_{i,2}, \ldots, n_{i,t-1}\}$ is the vector of annual past claims numbers observed at the beginning of each contract $t$. 

Considering each predictive distribution $N_{i,t}| \boldsymbol{n}_{i,(1:t-1)}$, we can calculate the frequency component premium of the contract $t$ ($t=1\ldots T$), denoted $\pi^{(F)}_{i,t}$, using the following equation: $\pi^{(F)}_{i,t} = E[N_{i,t}| \boldsymbol{n}_{i,(1:t-1)}]$. Therefore, the premium of contract $t$ of policy $i$ can be interpreted as a function of $\boldsymbol{n}_{(1:t-1)}$, which materializes the dependency between the $T$ contracts of insured $i$.

Usually, the classic actuarial approach to introducing dependency between the $T$ contracts of an insured $i$ is to introduce a random term familiar to all of the insured's $T$ contracts. See \citet{turcotte2022gamlss} or  \citet{pechon2019multivariate} for a review of this approach. Several other approaches in the literature have also been tried, such that of \citet{bermudez2018allowing} for models based on time series or \citet{shivaldez2014} for models using copula by the 'jittering' method.

More recently, instead of using the random effects approach, several papers have highlighted the advantages of using two families of models that take advantage of the fact that the average frequency of claims in property and casualty insurance is often between 0 and 1. Called Kappa-N model and Bonus-Malus Scales (BMS) model, these families propose to use a claims history function directly in the mean parameter of a counting distribution to model the decrease in premiums for insureds who do not file claims and the increase in premiums for insureds who do. Research on modelling the total annual claims number across several different databases and insurance products has shown that these models can generate excellent fit of training data and excellent prediction on test data, and often outperform the results of Bayesian random-effects models \citep{boucher2022,boucher2014}. 

In this paper, we generalize this BMS approach by working with data with a slightly more complex structure, close to what is used in practice, at least in North America. In Canada, some families contain multiple insured vehicles. From an experience-rating standpoint, the premium of one specific vehicle insured by a family could be calculated using the number of past claims for all others in this family, not just the number of past claims for that particular vehicle. In fact, in a family, the different drivers all use one or the other of the vehicles, so it makes sense to use the experience of all vehicles for rating. If we take as an example a family with two insured vehicles, we end up with a total premium defined as:

\begin{eqnarray*}
E[N_{i,1,t} + N_{i,2,t}| \boldsymbol{n}_{i,1(1:t-1)}, \boldsymbol{n}_{1,2,(1:t-1)}]
&=& 
E[N_{i,1,t}| \boldsymbol{n}_{(1:t-1)}, \boldsymbol{n}_{(1:t-1)}]
+
E[N_{i,1,t}| \boldsymbol{n}_{i,1,(1:t-1)}, \boldsymbol{n}_{i,2,(1:t-1)}] \\
&\ne& 
E[N_{i,1,t}| \boldsymbol{n}_{i,1,(1:t-1)}]
+
E[N_{i,2,t}| \boldsymbol{n}_{i,2(1:t-1)}] \ ,
\end{eqnarray*}

where $\boldsymbol{n}_{i,j,(1:t-1)} = \{n_{i,j,1}, n_{i,j,2}, \ldots, n_{i,j,t-1}\}$ is the vector of annual past claims numbers observed at the beginning of the contract $t$ of vehicle $j$ for this family $i$. 

Instead of counting the number of claims per vehicle in a family, we can count the number of claims per coverage/warranty \cite[see][for an illustration]{boucher2014}. We could also analyze the number of claims per insurance product, counting the number of home insurance claims to model the number of auto insurance claims. One can also consult \citep{verschuren2021predictive} for this type of application of BMS model. 

In this paper, we also propose to generalize the type of random variables modeled. Instead of using only the annual claims number for each contract $t$ of an insured $i$, we propose to develop a structure to model the cost of each claim $k$, $Z_{i,j,k,t}$ and  the annual claims amount $Y_{i,j,t}$. First, we deal with the joint distribution of the annual claims number and the costs of each these claims for the $T$ contracts of insured $i$. Second, we deal with the joint distribution of the annual claims number and the annual claims amount for these same $T$ contracts of insured $i$. The joint modelling of the annual claims number and the costs of each these claims is called frequency-severity modelling. See \citet{valdez2020,oh2020bonus,shi2019}, and \citet{shi2018pair} for a literature review about this model. Even if, in the loss cost model, the target variable is the annual claims amount of a contract, researchers recommend that this target variable and the annual claims number be modeled jointly \citep{delong2020,frees2014}.

\subsection{Terminologies and Definitions}
Similar to what has been done for the annual claims number modelling, one can also model the conditional distribution of the annual claims amount $Y_{i,j,t}$ according to the $t-1$ past annual claims amounts. However, in keeping with the idea of the Kappa-N and BMS models that are built conditionally on the number of past claims, our approach will be based on the analysis of the distribution of the annual claims amount, conditional on the number of past claims. In such a case, for an insured $i$, the premium of contract $t$ of vehicle $j$ is calculated as:

\begin{eqnarray*}
\pi^{(Y)}_{i,j,t} = E[Y_{i,j, t}|\boldsymbol{n}_{i, 1, (1:t-1)}, \ldots, \boldsymbol{n}_{i, J, (1:t-1)}] \ ,
\end{eqnarray*}

where $J$ is the total number of insured vehicles in the past. As shown in \ref{datastructure} and \ref{experiencerating}, the severity could also be modeled using this approach.

More generally, to cover all of these possibilities, \cite{boucher2022} defined two types of variables to be used in experience rating: 

\begin{enumerate}
\item The variable to model, named the \textbf{target variable};
\item The information used to define what we consider the past claim experience, named the \textbf{scope variable}.
\end{enumerate}

Using these definitions means that for this paper, three target variables will be modeled: the annual claims number (the claims frequency), the claims costs (the claims severity) and the annual claims amount (also called the loss cost). In contrast, all three target variables will be modeled based only on one type of scope variable: the number of past claims.  

\subsection{Summary}

In section \ref{datastructure} of the paper, we present some contextual elements and hypotheses to better introduce the models we propose. The models' notation is revised so that predictive pricing approaches can be used for the two new target variables, severity and loss cost. In the recent paper by \cite{delong2020}, the Compound Poisson-gamma (CPG) and Tweedie distributions were studied for their practical advantages in loss cost modelling. Our paper has the same objective as that of \citet{delong2020}, but we also consider the past insurance experience of each insured vehicle in calculating the premium of its future contracts. Details on how to use these two distribution in our proposed models is given in section \ref{experiencerating}. In section \ref{applicationnumerique}, we apply our proposed models to an auto insurance database of a major Canadian insurer. A variable selection step based on the \textit{Elastic-net} regularization is also introduced to measure the impact of adding a pricing component per experience. Section \ref{conclusion} concludes the paper.

\section{Data Structure and Hypotheses}\label{datastructure}

\subsection{Definitions and Form of Available Data}

We assume a hierarchical data structure in which the claims experience of the policies, vehicles and insurance contracts associated with each vehicle is observed. To ensure a common vocabulary, we have retained some of the terms used in the introduction but have clarified their definitions further:

\begin{itemize}
    \item A \textbf{policy} is usually associated with a single insured. In insurers’ databases, a policy is usually identified by
    a unique number. 
    \item An insured vehicle, or simply a \textbf{vehicle}, is associated with a policy. For an in-force policy, the minimum number of insured vehicles is one, but many insured (particularly in North America) might have several vehicles.
    \item A policy is often made up of several insurance \textbf{contracts}. An insurance contract is also often referred
    to as an insurance term, and it is usually one year long. Insurance contracts are sometimes shorter, for example three
    or six months. Some insurance policies contain only one term, but a significant portion of policies contain multiple
    contracts. 
\end{itemize}

For a given policy $i, i=1, \ldots, m$, we assume that the claims of a $j, j=1 \ldots, J_i$ vehicle are observable through $T_{i,j}$ contracts. We denote by $t$, the index associated with the $T$ contracts ($i$ and $j$ are removed to simplify reading). Our variables of interest are therefore the claims experience of the $T$ contracts of each vehicle in a policy. 

To better capture the form of the available data, Table \ref{Firstillustrationbase} provides an illustration of a sample of three insurance policies.  It shows that policy \#1 contained only one insured vehicle for the 2018 and 2021 policies, but two vehicles were insured in 2019 and 2020.  Thus, the claims experience of the first vehicle in policy \#1 was observed during four annual contracts, while the claims experience of the second vehicle was observed during only two annual contracts. Policy \#2 in this sample contains only one insured vehicle, and that vehicle was insured for only one annual contract. Finally, two vehicles insured on a single annual contract were observed in the third and final policy in this table. 

It should be noted that the contract number is obtained according to its associated policy and its effective date. That is, in a given policy, all contracts with the same effective year have the same contract number. The first contract for vehicle \#2 of policy \#1 illustrates this situation. Such a notation is important for the rest of the paper. 

As can be seen in the sample in the same table, the characteristics of the insured and the insured vehicle are also available.  Finally, the frequency of claims and the cost of each claim are also available information. The loss cost, representing the sum of the costs of each claim, is shown in the last column of the table. An insured who has not made any claims during their contract execution period has a loss cost of zero. 

\begin{table}[ht]
	\begin{center}
		\begin{adjustbox}{max width=\textwidth}
			\begin{tabular}{|c|c|c|c|c|c|c|c|c|c|c|c|}
				\hline
				Policy & Vehicle & Contract & Contract &  \multicolumn{3}{|c|}{Risk Characteristics} & Number of & \multicolumn{3}{|c|}{Costs of Claim}  & Loss\\ 
				Number & Number & Number & Date & Age & Sex & $\ldots$ & Claims & \#1 & \#2 & \#3 & Costs\\ 
				\hline
				1 & 1 & 1 & 2018-01-15 & 42 & M & $\ldots$ &  0 & . & .& .& 0\\ 
				1 & 1 & 2 & 2019-01-15 & 43 & M & $\ldots$ &  2 & 6,592 & 11,520& .&18,112\\ 
				1 & 1 & 3 & 2020-01-15 & 44 & M & $\ldots$ &  1 & 24,151 & .& .&24,151\\ 
				1 & 1 & 4 & 2021-01-15 & 45 & M & $\ldots$ &  0 & . & .& .& 0\\ 
				1 & 2 & 2 & 2019-01-15 & 40 & F & $\ldots$ &  2 & 1,490 & 24,505 & .&25,995 \\ 
				1 & 2 & 3 & 2020-01-15 & 41 & F & $\ldots$ &  0 & . & . & .&0\\ 
				2 & 1 & 1 & 2018-02-05 & 24 & M & $\ldots$  &  0 & . & .& .&0\\
    		3 & 1 & 1 & 2018-02-08 & 34 & F & $\ldots$  &  0 & . & .& .&0\\
          	3 & 2 & 1 & 2018-02-08 & 30 & M & $\ldots$  &  1 & 8,150 & .& .&8,150\\
				\hline
			\end{tabular}
		\end{adjustbox}
            \caption{Illustration of frequency and severity data }
            \label{Firstillustrationbase}
	\end{center}
\end{table}

\subsection{Target Variables}

For vehicle $j$ of an insured $i$, the random variable $N_{i,j,t}$ represents the annual claims number of its contract $t$. If the observed annual claims number $n_{i,j,t} = n > 0$, the random vector $Z_{i,j,t}=\left(Z_{i,j,t,1},...,Z_{i,j,t, n}\right)^{'}$ will represent the vector of each the insured's $n$ claims costs. This is not defined if the associated observed annual claims number $n_{i,j,t}= n = 0$. 

Thus, we calculate the loss cost, denoted by the random variable $Y_{i,j,t}$ as follows:

\[Y_{i,j,t}=
\begin{cases} 
\sum_{k=1}^{N_{i,j,t}}Z_{i,j,t,k}&\text{if  $N_{i,j,t} >0$}\\ \\
	0 & \text{if $N_{i,j,t}=0$}
\end{cases}= \left(\sum_{k=1}^{N_{i,j,t}}Z_{i,j,t,k}  \right)\indic\left(N_{i,j,t} >0 \right).
\] 

We define our three target variables by the following three random variables: $N_{i,j,t}$, $Z_{i,j,t}$ and $Y_{i,j,t}$. 

\subsubsection{Premiums}
For the $m$ insured in the portfolio, assuming a minimization of the square distance for the calculation of the premium of contract $t$ for each vehicle $j$  in the policy $i$, the parameter of interest corresponds to $\Esp{Y_{i,j,t}},~ \forall i= 1,...,m,~ \forall j= 1,\ldots,J_i,~ \forall t = 1,\ldots,T_{i,j}$. This parameter can be calculated in two ways: (1) By the multiplying the frequency component premium and the severity component premium according to some assumptions; (2) By considering the conditional distribution of the loss cost denoted by $f_{Y_{i,j,t}}(.)$. Formally, these two ways are expressed as:

\[	
 \Esp{Y_{i,j,t}}=
\begin{cases} 
 \Esp{N_{i,j,t}} \Esp{Z_{i,j,t,k}}  & \text{ (1)}\\ \\
\int y f_{Y_{i,j,t}}(y) \, dy  & \text{(2)}.
\end{cases}
\] 

The assumptions to obtain a premium according to (1) are generally defined as follows:

\begin{enumerate}
   \item The independence between the claims frequency and the claims severity for the same contract $t$: 
   $$N_{i,j,t} \indep Z_{i,j,t,k}, ~ \forall i= 1,\ldots,m,~ \forall j= 1,\ldots,J_i,~ \forall t = 1,\ldots,T_{i,j}, ~ \forall k = 1,\ldots, n_{i,j,t}.$$
  \item The independence of the costs of each claim across distinct policies: $$Z_{i,j,t,k_1}  \indep  Z_{i,j,t,j,k_2},
	~ \forall i= 1,\ldots,m,~ \forall j= 1,\ldots,J_i,~ \forall t = 1,...,T_{i,j}, k_1 \ne k_2.$$
   \item For the same contract $t$, the costs of each claim are identically distributed.

\end{enumerate}

In order to include some form of segmentation in the rating \citep{frees2014}, it should be noted that the premium is generally calculated considering specific observable characteristics of each contract, such as those illustrated in Table \ref{Firstillustrationbase}. We denote these characteristics by the following vector $\boldsymbol{X}_{i,j,t} =\left(x_{i,j,t, 0},...,x_{i,j,t,q}\right)^{'} \in \mathbb{X} \subset \{1\}\times \mathbb{R}^q, q>0$. We are finally interested in these quantities: $\pi_{i,j,t}^{(Y)} =\Esp{Y_{i,j,t}|\boldsymbol{X}_{i,j,t}}$, $\pi_{i,j,t}^{(N)} =\Esp{N_{i,j,t}|\boldsymbol{X}_{i,j,t}}$, $\pi_{i,j,t}^{(Z)} =\Esp{Z_{i,j,t}|\boldsymbol{X}_{i,j,t}}$.

\section{Experience Rating with Compound Poisson-gamma (CPG) and Tweedie Models} \label{experiencerating}

For the experience rating, the Kappa-N and BMS models are generally proposed \citep{boucher2022}, which model the conditional distribution of a target variable according to the scope variables. In this section, the CPG and Tweedie are used as an underlying distribution in each model. Before presenting Kappa-N and BMS models in our context, we start with an example to  better explain how the scope variables are calculated in practice. 

\subsection{Scope variables}

It is known in actuarial science that insureds who make claims will have a higher frequency of claims in their future contracts. This can be explained in several ways: some insureds behave more riskily than others, some insureds live in areas that are more prone to disasters, and some insured property is more likely to be damaged. Individual characteristics used as segmentation variables may partly explain this situation. However, many of these variables cannot be measured and modeled directly in rating. Thus, past claims experience can be used to approximate the effect of these non-measurable characteristics on premiums. This is why, in addition to conditioning on characteristics $\boldsymbol{X}_{i,j,t}$, we price an insured according to their claims history, defined as a  scope variable in the introduction.

To illustrate the situation adequately, we use Table \ref{Firstillustrationbase} as an example, which we generalize to Table \ref{Secondillustrationbase}. For each vehicle in Table \ref{Firstillustrationbase}, we can calculate the number of claims from past contracts, i.e $\boldsymbol{n}_{i,j,(1:t-1)} = \{n_{i,j,1}, n_{i,j,2}, \ldots, n_{i,j,t-1}\}$. This is shown in columns 5, 6 and 7 of Table \ref{Secondillustrationbase}. However, our scope  variable of  frequency will not only be composed of the number of  past claims for the same vehicle, but also the number of past claims of the entire policy.  Thus, in the last three columns of Table \ref{Secondillustrationbase}, the sum of the  past claims for all vehicles of the same policy is shown for the previous contracts, namely:

\begin{align*}
\boldsymbol{n}_{i,\bullet,(1:t-1)} &= \left\{\sum_{j=1}^{J_i} n_{i,j,1}, \sum_{j=1}^{J_i} n_{i,j,2}, \ldots, \sum_{j=1}^{J_i} n_{i,j,t-1}\right\} 
= \{n_{i,\bullet,1}, n_{i,\bullet,2}, \ldots, n_{i,\bullet,t-1}\}.
\end{align*}

\begin{table}[ht]
\begin{center}
\begin{adjustbox}{max width=\textwidth}
    \begin{tabular}{|c|c|c|c|c|c|c|c|c|c|}
        \hline
        Policy & Vehicle & Contract &  & \multicolumn{6}{|c|}{Scope Variables} \\
        $i$ & $j$ & $t$ & $n_{i,j,t}$ & $n_{i,j,t-1}$ & $n_{i,j,t-2}$ & $n_{i,j,t-3}$ & $n_{i,\bullet,t-1}$ & $n_{i,\bullet,t-2}$ & $n_{i,\bullet,t-3}$ \\ 
        \hline
        1 & 1 & 1 & 0 & . & . & . &  .  & . & .  \\ 
        1 & 1 & 2 & 2 & 0 & . & 0 &  0  & . & .  \\ 
        1 & 1 & 3 & 1 & 2 & 0 & 2 &  4  & 0 & .  \\ 
        1 & 1 & 4 & 0 & 1 & 2 & 0 &  1  & 4 & 0  \\ 
        1 & 2 & 2 & 2 & . & . & 0 &  0  & . & .  \\ 
        1 & 2 & 3 & 0 & 2 & . & . &  4  & 0 & .  \\ 
        2 & 1 & 1 & 0 & . & . & . &  .  & . & .  \\
        3 & 1 & 1 & 0 & . & . & . &  .  & . & .  \\
        3 & 2 & 1 & 1 & 0 & . & . &  1  & 0 & .  \\
        \hline
    \end{tabular}
\end{adjustbox}
        \caption{Illustration of scope variables}
        \label{Secondillustrationbase}
\end{center}
\end{table}

\subsection{Kappa-N Models}
\label{KappaNSection}

For a loss cost model, we are looking for the joint conditional distribution of $(N_{i,j,t},Y_{i,j,t})$ according to $\boldsymbol{n}_{i,\bullet,(1:t-1)}$ and $\boldsymbol{X}_{i,j,t}$. For a frequency-severity model, we are looking for the joint conditional distribution of $(N_{i,j,t},  Z_{i,j,t})$ according to $\boldsymbol{n}_{i,\bullet,(1:t-1)}$ and $\boldsymbol{X}_{j,t}$.

Using a logarithmic link between the co-variables and the mean parameter such as in GLM models \citep{delong2020,jong}, the expectations for our three variables of interest are expressed as given by Equations (\ref{expec1}), (\ref{expec2}) and (\ref{expec3}) where $h^{(Y)}(.)$, $h^{(N)}(.)$ et $ h^{(Z)}(.)$ represent the functions of historical claims. 
 
\begin{align} \label{expec1}
 	\pi_{i,j,t}^{(N)}  &= \exp\left(\boldsymbol{X}^{'}_{i,j,t} \beta^{(N)} + h^{(N)}(n_{i,\bullet, 1}, \ldots, n_{i,\bullet, t-1})\right),~ \beta^{(N)} = (\beta^{(N)}_0,\ldots,\beta^{(N)}_q)^{'} \in \mathbb{R}^{q + 1}. \\\label{expec2}
 	\pi_{i,j,t}^{(Z)}  &= \exp\left(\boldsymbol{X}^{'}_{i,j,t} \beta^{(Z)} + h^{(Z)}(n_{i,\bullet, 1}, \ldots, n_{i,\bullet, t-1})\right),~ \beta^{(Z)} = (\beta^{(Z)}_0,\ldots,\beta^{(Z)}_q)^{'} \in \mathbb{R}^{q + 1}. \\\label{expec3}
   	\pi_{i,j,t}^{(Y)}  &= \exp\left(\boldsymbol{X}^{'}_{i,j,t} \beta^{(Y)} + h^{(Y)}(n_{i,\bullet, 1}, \ldots, n_{i,\bullet, t-1})\right),~ \beta^{(Y)} = (\beta^{(Y)}_0,...,\beta^{(Y)}_q)^{'} \in \mathbb{R}^{q + 1}.  
\end{align}

It should be noted that several possibilities exist to define these historical claims functions. \cite{boucher2022} listed some of them, and the problems that they could create. Taking advantage of the fact that the average automobile insurance claim frequency is between 0 and 1, and that insureds expect a discount when they do not claim and a surcharge if they report a claim, Boucher proposed instead to define a new indicator variable  $\kappa_{i,j,t} = I(n_{i,j,t} = 0)$ that identifies claims-free contracts. We thus have two new variables summarizing the claims experience:

\begin{align*}
n_{i,\bullet, \bullet} &= \sum_{\tau=1}^{t-1} n_{i,\bullet, \tau} \ \text{, and}  \ \ 
\kappa_{i,\bullet, \bullet} = \sum_{\tau=1}^{t-1} \kappa_{i,\bullet, \tau} = \sum_{\tau=1}^{t-1} I(n_{i,\bullet, \tau}=0)\ , \ \text{ and so}  \ \ \\
 &h^{(.)}(n_{i,\bullet, 1}, \ldots, n_{i,\bullet, t-1}) = -\gamma_0^{(.)} \kappa_{i,\bullet, \bullet} + \gamma_1^{(.)} n_{i,\bullet, \bullet} \ ,
\end{align*}

where $\gamma_0^{(.)}, \gamma_1^{(.)} \in \mathbb{R}$. The negative sign in front of the positive parameter $\gamma^{(.)}_0$ is used to highlight that an additional year without a claim will decrease the premium. This simple way of summarizing the claims history in the mean parameter of a random variable is called Kappa-N modelling. The idea is to consider $\kappa_{i,\bullet, \bullet}$ and $n_{i,\bullet, \bullet}$ as co-variables in premium modelling.

\subsubsection{Claims Score}
For each policy $i$ and each vehicle's contract $t$, we define a positive quantity $\ell_{i,t}^{(.)}$ called the claims score based on the function $h^{(.)}(.)$ and an initial score $\ell_1$. This initial score is interpreted as the maximum number of years for which a contract can remain without a claim from its effective date to its end date. \cite{boucher2022} sets $\ell_1 = 100$ for a simple aesthetic reason:

\begin{align*}
\ell_{i,t}^{(.)} &= \ell_{1} + \frac{1}{\gamma_0^{(.)}} h^{(.)}(n_{i,\bullet, 1}, \ldots, n_{i,\bullet, t-1}) \\
&=\ell_{1} + \frac{1}{\gamma_0^{(.)}} \left( -\gamma_0^{(.)} \kappa_{i,\bullet, \bullet} + \gamma_1^{(.)} n_{i,\bullet, \bullet} \right) \\
&= \ell_{1} - \kappa_{i, \bullet, \bullet} + \Psi^{(.)} n_{i, \bullet, \bullet}, ~ \text{ with }\Psi^{(.)} = \frac{\gamma_1^{(.)}}{\gamma_0^{(.)}} \ ,
\end{align*}

where $\Psi^{(.)}$ is the jump parameter and $\gamma_0$ is the relativity parameter of penalties related to past claims. 

For a Kappa-N model using the claims score, the expectations of our three variables of interest can be expressed as:

\begin{align}
\pi_{i,j,t}^{(N)}  &= \exp\left(\boldsymbol{X}^{'}_{i,j,t} \beta^{(N)} + \gamma_0^{(N)} \ell_{i,t}^{(N)} \right).\\
\pi_{i,j,t}^{(Z)}  &= \exp\left(\boldsymbol{X}^{'}_{i,j,t} \beta^{(Z)} +  \gamma_0^{(Z)} \ell_{i,t}^{(Z)} \right).\\
\pi_{i,j,t}^{(Y)}  &= \exp\left(\boldsymbol{X}^{'}_{i,j,t} \beta^{(Y)} + \gamma_0^{(Y)} \ell_{i,i,t}^{(Y)} \right). 
\end{align}

From these equations, one can quickly assess the impact of the claims score on the insurance premium. This has several desirable qualities in terms of the contract's rating structure:  

\begin{itemize}
    \item For an insured $i$ without insurance experience,  $n_{i, \bullet, \bullet}=0$ and $\kappa_{i, \bullet, \bullet}=0$, which means a claim score of $\ell_t = 100 = \ell_1$; 
    \item Each annual contract without a claim will decrease the claim score $\ell$ by 1;
    \item Each claim increases the claim score $\ell$ by $\Psi$;
    \item The impact of a single claim on the premium is then roughly equal to $\Psi$ years without claims;
    \item The penalty for a claim is an increase of $(\exp(\Psi \gamma_0) - 1)$\% of the premium;
    \item Each year without a claim decreases the premium by $(1 - \exp(-\gamma_0))$\%.
\end{itemize}

\subsubsection{Kappa-N For Independent Poisson Annual Claims Numbers }

Considering the risk exposure of contract $t$, denoted by $d_{i,j,t}$, we assume that its annual claims number $N_{i,j,t} | \ell_{i,t}^{(N)}$ is Poisson distributed, such that:

\begin{align}\label{primefrequence}
	\pi_{i,j,t}^{(N)} &=  d_{i,j,t}\exp\left(\boldsymbol{X}^{'}_{i,j,t} \beta^{(N)} + \gamma_0^{(N)} \ell_{i,t}^{(N)} \right) \ , \\\label{scorepoisson}
	\ell_{i,t}^{(N)}&=\ell_{1} - \kappa_{i, \bullet, \bullet} + \Psi^{(N)} n_{i, \bullet, \bullet}, ~ \text{ with }\Psi^{(N)} = \frac{\gamma_1^{(N)}}{\gamma_0^{(N)}}.
\end{align}

For inference  purposes , we assume the independence  between the contracts' annual claims number of distinct policies:

$$N_{i_1, j, t} \indep N_{i_2, j, t},~\forall i_1,i_2 = 1,...,n| i_1\ne i_2.$$

Further, given the use of the claims score $\ell^{(N)}$, a form of dependency will exist between contracts for the same vehicle, and contracts for vehicles of the same policy.  More formally, we have:

$$N_{i, j_1, t_1} \nindep N_{i, j_2, t_2}, ,~\forall j_1,j_2,t_1,t_2.$$

The likelihood contribution for the claims frequency of a single policy  $i$ is expressed as follows ($i$ is removed for easy reading):

\begin{equation} \label{contributionlogpois}
	\prod_{j=1}^{J} \prod_{t=1}^{T} \exp\left( - \pi_{j,t}^{(N)}  +  n_{j,t} \log\left(  \pi_{j,t}^{(N)}\right) - \log\left( n_{j,t}!\right)\right).
\end{equation}

Finally, the idea is to estimate the parameters $\beta^{(N)}$, $\gamma_0^{(N)}$ and $\gamma_1^{(N)}$ by maximizing the likelihood function built by multiplying the contribution (\ref{contributionlogpois}) for $m$ policies. This optimization can also be done using the \textit{glm} function in R.  
 
\subsubsection{Kappa-N for Independent Gamma Claims Costs }
For a contract $t$, we assume that the common distribution of the costs of each of its $n_{i,j,t} = n$ claims, $Z_{i,j,t,k}$, is gamma such that:  

\begin{align}\label{primeseverite}
	\pi_{i,j,t}^{(Z)} & = \exp\left(\boldsymbol{X}^{'}_{i,j,t} \beta^{(Z)} +  \gamma_0^{(Z)} \ell_{i,t}^{(Z)} \right) \ ,\\\label{scoregamma}
	\ell_{i,t}^{(Z)}&=\ell_{1} - \kappa_{i, \bullet, \bullet} + \Psi^{(Z)} n_{i, \bullet, \bullet}, ~ \text{ with }\Psi^{(Z)} = \frac{\gamma_1^{(Z)}}{\gamma_0^{(Z)}}.
\end{align}

It is important to note that the cost of a claim, $Z_{i,j,t,k}, k=1, \ldots, n$ depends on the score  $\ell^{(Z)}$ which is set at the beginning of period $t$. Thus, the first, second, and third claims of contract t are all dependent on the same score $\ell^{(Z)}$. This score, as expressed in equation (\ref{scoregamma}), will only be updated at the end of contract $t$ for the rating of contract $t+1$.

Similar to the frequency part, we assume that the claims severities of contracts of distinct policies are independent. Further, we take into account the dependency between the severity of contracts for the same vehicle and between those of vehicles of the same policy:  

\begin{align*}
	Z_{i_1, j, t, k} &\indep Z_{i_2, j, t, k},~\forall i_1,i_2 = 1,...,n| i_1\ne i_2. \\
	Z_{i, j_1, t_1, k} &\nindep Z_{i, j_2, t_2, k}, ,~\forall j_1,j_2,t_1,t_2.
\end{align*}

We therefore evaluate the  contribution  of the likelihood for the severity of a policy $i$ as follows ($i$ is removed for easy reading): 

\begin{equation} \label{contributionloggam}
	\prod_{j=1}^{J} \prod_{t=1}^{T}\prod_{k=1}^{n} \exp\left(  - \frac{\gamma}{ \pi_{j,t}^{(Z)}} z_{j,t,k} - \gamma \log\left( \pi_{j,t}^{(Z)}\right)  - \log\left( \frac{z^{\gamma - 1}_{j,t,k} \gamma^{\gamma}}{\Gamma(\gamma)}   \right)\right) \ ,
\end{equation}

where $z_{j,t,k}$, is the observed cost of claim $k$ of contract $t$. Thus, the inference consists in maximizing the likelihood function built from the likelihood contributions (\ref{contributionloggam}) of the $m$ policies. Similar to the Poisson model, the \textit{glm} function in R can also be used to estimate the parameters $\beta^{(Z)}$, $\gamma_0^{(Z)}$ and $\gamma_1^{(Z)}$.

\subsubsection{Remarks}

In the \textbf{Compound Poisson-gamma (CPG)} model, for each contract, it is assumed that the individual's annual claims number is Poisson distributed and the costs of each of the insudred's claims is gamma distributed. This model also assumes independence between the frequency and the severity of claims of each contract. Therefore, to calculate the annual premium of a contract, one can model  its frequency and severity components separately \citep{delong2020}. 

For the severity modelling, we consider the costs of each claim, unlike \cite{delong2020}, who considered the average claims amount of each contract as a target variable. Note that these two approaches lead to the same inference results \citep{delong2020}.

\subsubsection{Kappa-N for Independent Tweedie Annual Claims Amount }

Instead of using the CPG approach to model the loss cost of a contract, another alternative is to use the distribution of its annual claims amount directly and calculate the expectation of this distribution to obtain the annual premium. This is the idea of the Tweedie model. 

Consistent with \citet{delong2020}, we consider for each contract $t$, the couple of random variables $(N_{i,j,t}, Y_{i,j,t})$ representing the annual claims number and the annual claims cost: $Y_{i,j,t}$ is Tweedie distributed and $N_{i,j,t}$ is Poisson distributed. 

For inference purposes, we are interested in the conditional distribution of $(N_{i,j,t}, Y_{i,j,t})$ according to the $\ell_{i,t}^{(Y)}$ and $\boldsymbol{X}_{i,j,t}$. We also assume the following equations for the  mean $\mu_{i,j,t}$ and the dispersion  $\phi_{i,j,t}$ parameters of the Tweedie distribution: 

\begin{align}
	\pi^{(Y)}_{i,j,t} &= d_{i,j,t}\exp\left(\boldsymbol{X}^{'}_{i,j,t} \beta^{(Y)} + \gamma_0^{(Y)} \ell^{(Y)}_{i,t} \right) = \mu_{i,j,t} \ ,\\
	\phi_{i,j,t} &= \exp\left( \bm{X^{'}_{i,j,t}} \beta^{(D)} + \gamma^{(D)}_0 \ell^{(Y)}_{i,t}\right) \ ,\\
	\label{scoretweedie}
    \ell_{i,t}^{(Y)}&=\ell_{1} - \kappa_{i, \bullet, \bullet} + \Psi^{(Y)} n_{i, \bullet, \bullet}, ~ \text{ with }\Psi^{(Y)} = \frac{\gamma_1^{(Y)}}{\gamma_0^{(Y)}} \ ,
\end{align}

where $\beta^{(D)}$ is a real vector of the same dimension as $\beta^{(Y)}$ and $\gamma^{(D)}_0$ is a real parameter.  

According to \citet{delong2020}, we can also obtain the probability density function of $(N_{i,j,t}, Y_{i,j,t})$ as follows ($i$ is removed for easy reading):

\[ f_{j,t}(y,n) = \begin{cases} 
	\prod_{t=1}^{T} \exp \left\{ \frac{w_{j,t}}{\phi_{j,t}}\left(y\frac{\mu^{1 - p}_{j,t}}{1 - p} - \frac{\mu^{2 - p}_{j,t}}{2 - p}  \right) + \log\left(  \frac{\left( \left( \frac{w_{j,t}}{\phi_{j,t}}\right)^{\gamma +1 } y^{\gamma} \right)^{n}}{n!\Gamma(n \gamma) y  \left(p - 1\right)^{\gamma n}\left(2 - p\right)^{n} }  \right) \right 
	\}& \text{$n > 0$}\\\\
	\prod_{t=1}^{T_{i,j}} \exp\left\{ - \frac{w_{j,t}\mu^{2 - p}_{j,t}}{(2 - p)\phi_{j,t}} \right\}&
	\text{$n = 0$} \ ,
\end{cases}
\]

where $w_{j,t}$ and $p$ represent respectively the weight and the Tweedie variance parameter. $p$ is a positive real and verifies: $1< p < 2$. The parameter $\gamma$ is obtained as: $\gamma = \frac{2 - p}{p - 1}$. \\  

By assuming the independence between the loss costs of contracts of distinct policies, we can calculate the contribution of policy $i$ to the likelihood as ($i$ is removed for easy reading): 

\begin{equation}\label{contributionlogtwe}
	\prod_{j=1}^{J} \prod_{t=1}^{T} f_{j,t}\left(y_{j,t},n_{j,t}\right).
\end{equation}

To estimate all parameters, one can use the maximum likelihood strategy considering the (\ref{contributionlogtwe}). The idea is to define for each policy $i$, each vehicle $j$ and each contract $t$, a response variable for the dispersion parameter $\phi_{i,j,t}$ as follows:  

$$D_{i,j,t}= \frac{2}{\nu_{i,j,t}}\left( -w_{i,j,t}\left(Y_{i,j,t} \frac{\mu^{1 -p}_{i,j,t}}{1 - p} - \frac{\mu^{2 -p}_{i,j,t}}{2 - p}\right) - \phi_{i,j,t}\frac{N_{i,j,t}}{p - 1} \right) + \phi_{i,j,t} \ ,$$

where  $\nu_{i,j,t} = \frac{2w_{i,j,t}}{\phi_{i,j,t}}\frac{\mu^{2 -p}_{i,j,t}}{(p - 1)(2 -p)}.$

The main motivation of this response variable is that we obtain: $\Esp{D_{i,j,t}} = \phi_{i,j,t}$. Therefore, the optimization of the likelihood function can be seen as two connected GLM (Generalized Linear Models): 1) GlM for the mean parameter; and 2) GLM for the dispersion parameter. This approach is called Double-GLM (DGLM) \citep{delong2020}. 

\subsection{Remarks} \label{remarques}
We close this section on Kappa-N models with a few remarks about the underlying distributions used.

\subsubsection{Tweedie Case}
For practical reasons, in our analysis, we carefully distinguish between the risk exposure $d_{i,j,t}$  of a contract and the  weight  $w_{i,j,t}$ of the Tweedie distribution. As previously noted, $d_{i,j,t}$ is a correction factor for the annual premium, whereas weight is a parameter that influences the Tweedie distribution modelling. In our analysis, we find that some values of the weight lead to an overestimation of the amount of total losses. For this reason, we suggest to consider the weight  $w_{i,j,t} =  d^{p-1}_{i,j,t}$. However, there will always be a difference between the total amount of losses and the total amount of expected losses. Even so, the difference between these two quantities is not very large. To correct this, one can adopt the off-balance correction as mentioned in the paper by \citep{denuit2021autocalibration}.

\subsubsection{Tweedie vs CPG }
By considering the log-likelihood criterion, \cite{delong2020} showed that the Tweedie model is preferable to the CPG model. However,  to use this log-likelihood criterion, the data representation should be the same in each model. \citet{delong2020} proposed a theorem (\textit{Theorem 3.8} in their paper) to compare the log-likelihood obtained in each model. We consider the same theorem in our analysis ($i$ is removed for easy reading):
\begin{align} \label{cpmoyenne}
	\bm{\tilde{X}^{'}_{j,t}}\beta^{*Y} &= \bm{X^{'}_{j,t} }\beta^{N} +   \bm{X^{'}_{j,t}} \beta^{Z}  + \gamma_0^{(N)}\ell^{(N)}_{t} + \gamma_0^{(Z)} \ell^{(Z)}_{t} \ ,\\\label{cpdispersion}
	\bm{\tilde{X}^{'}_{j,t} }\beta^{*D}&= - \log(2 - p) - (p - 1)\bm{X^{'}_{j,t}}\beta^{N} + (2 - p) \bm{X^{'}_{j,t}} \beta^{Z} - (p - 1)\gamma_0^{(N)}\ell^{(N)}_{t} + (2 - p)\gamma_0^{(Z)} \ell^{(Z)}_{t} \ ,
\end{align}

where $\tilde{X}_{j,t} = \left(x_{j,t, 0},...,x_{j,t,q}, \ell^{(N)}_{t}, \ell^{(Z)}_{t}\right)^{'}$, $\beta^{*Y}$ and $\beta^{*D}$ are respectively the mean and the dispersion parameters of CPG in Tweedie parametrization. 

Even if  $\tilde{X}_{j,t} \ne \left(x_{j,t, 0},...,x_{j,t,q}, \ell^{(Y)}_{t}\right)^{'}$, we are able to compare the two models using (\ref{cpmoyenne}) and (\ref{cpdispersion}).

\subsection{Bonus-Malus Scale Models}
\label{BMSSection}

One of the practical problems with the Kappa-N model is the excessive increase and decrease in premiums due to annual penalties and discounts. In order to limit the variation of premiums over time and to allow some form of forgiveness of old claims in the rating, another approach constrains the score  $\ell$ to be between two limits for all the past contracts. \cite{boucher2022} called this approach the Bonus-Malus Scale (BMS) Models. See \citet{lemaire} and \citet{denuit2007} for a historical review of BMS models.  
Instead of interpreting $\ell_{i,t}$ as a claims score, it can simply mean the BMS level. For an insured $i$, we define this BMS level as:

\begin{align}
	\ell_{i,t} = \ell_{i,t-1} -  \sum_{j=1}^{J_i} \kappa_{i,j,t -1} + \Psi \times  \sum_{j=1}^{J_i} n_{i,j,t-1}, \text{ with } \ell_{min} \le \ell_{i,t} \le \ell_{max}, \forall t = 1,\ldots,T. \label{simpletransition}
\end{align}
	
Recursively, for any policy $i$ and any contract $t$, the BMS level, $\ell_{i,t}$, is obtained as follows:

\begin{align*}
\ell_{i,t} & = \ell_{i,1} - \sum_{\tau = 1}^{t-1} \sum_{j=1}^{J_i} \kappa_{i,j,\tau -1} + \psi  \sum_{\tau = 1}^{t-1}  \sum_{j=1}^{J_i} n_{i,j,\tau-1} = \ell_{i,1} - \kappa_{i, \bullet, \bullet} + \psi n_{i, \bullet, \bullet}.
\end{align*}

It should be noted that these recursive equations are true regardless of the $\ell_{min}$ and $\ell_{max}$ limits, and confer the Markov property on the Kappa-N and BMS models. The starting BMS level $\ell_1$ is set at $100$ as in the Kappa-N model. Thus, for our three variables of interest, the corresponding premiums are calculated using Table \ref{bmspremiums}. We note that the difficulty of the inference is the definition of the BMS level, which is a co-variable and an endogenous variable to the model simultaneously. This BMS level depends on the parameters $\Psi$, $\ell_{min}$ and $\ell_{max}$ which are all unknown in the model. However, for the inference, one can use the profile maximization of the likelihood function on the three parameters: $\Psi$, $\ell_{min}$ and $\ell_{max}$. The idea is to use all possible values of these three parameters to estimate the other parameters of the model. For the jump parameter, $\Psi$, the use of natural numbers is required.

\begin{table}[ht]
	\begin{center}
		\begin{adjustbox}{max width=\textwidth}
			\begin{tabular}{|c|c|c|}
				\hline
				 Model & Premium &BMS level\\  
				\hline
 &&\\				
 & & $\ell^{(N)}_{i,t} =\ell_1 - \kappa_{i, \bullet, \bullet} + \Psi^{(N)} n_{i, \bullet, \bullet}$ \\ 
Frequency&$\pi^{(N)}_{i,j,t} = d_{i,j,t}\exp\left(\boldsymbol{X}^{'}_{i,j,t} \beta^{(N)} + \gamma_0^{(N)} \ell^{(N)}_{i,t} \right)$ &\\
&&$ \ell^{(N)}_{min} \leq \ell^{(N)}_{it} \leq \ell^{(N)}_{max} $\\
 &&\\
	\hline
	 &&\\
 &  & $\ell^{(Z)}_{i,t} =\ell_1 - \kappa_{i, \bullet, \bullet} + \Psi^{(Z)} n_{i, \bullet, \bullet}$\\
Severity&$\pi^{(Z)}_{i,j,t} = \exp\left(\boldsymbol{X}^{'}_{i,j,t} \beta^{(N)} + \gamma_0^{(Z)} \ell^{(Z)}_{i,t} \right)$&\\
&&$ \ell^{(Z)}_{min} \leq \ell^{(Z)}_{it} \leq \ell^{(Z)}_{max}$\\ 
 &&\\
	\hline
	 &&\\
& & $\ell^{(Y)}_{i,t} =\ell_1 - \kappa_{i, \bullet, \bullet} + \Psi^{(Y)} n_{i, \bullet, \bullet}$ \\ 
Loss Cost &$\pi^{(Y)}_{i,j,t} = d_{i,j,t}\exp\left(\boldsymbol{X}^{'}_{i,j,t} \beta^{(Y)} + \gamma_0^{(Y)} \ell^{(Y)}_{i,t} \right)$&\\
&&$ \ell^{(Y)}_{min} \leq \ell^{(Y)}_{it} \leq \ell^{(Y)}_{max} $\\
 &&\\
				\hline
			\end{tabular}
		\end{adjustbox}
		\caption{Premiums in BMS models }
		\label{bmspremiums}
	\end{center}
\end{table}

\section{Numerical application}\label{applicationnumerique}

\subsection{Description of data} 
We consider a non-random sample of an automobile insurance database of a major insurer in Canada over a total period of 13 consecutive years. The data concern the Canadian province of Ontario and contain more than 2 million observations. Each observation corresponds to an annual contract for one vehicle. The form of the database is similar to Table \ref{Firstillustrationbase} introduced at the beginning of our paper. For each observation in the database, we have a policy number, a vehicle number as well as the effective and end date of the vehicle contract. Several characteristics of the insured or the insured vehicle are also available.  Finally, for each contract for each vehicle, the number of claims and the cost of each claim are available.

The database is also divided into coverage type, which provides information on third-party liability, collision and comprehensive claims. To illustrate the approach described in this paper, we focused on a single cover. Thus, in connection with the  defined terms in the introduction, we illustrate our pricing model by experience with:

\begin{itemize}
	\item A target variable based on collision coverage, representing the property damage protection of auto insurance for accidents for which the driver is at fault; 
	\item  A scope variable also based on collision coverage. 	
\end{itemize}

As mentioned earlier, however, the proposed pricing approach is very flexible, and any combination of target and scope variables would be possible. For example, the analysis of the sum of liability and collision coverage could be analyzed by conditioning on the experience of past claims of comprehensive coverage.

For confidentiality reasons, the full  description of the data will not be detailed. That being said, we can provide some summary information for the studied sample:

\begin{itemize}
	\item The observed annual claims frequency is approximately 2\%;
	\item The average severity of a claim is around \$7,500; 
	\item The average annual loss cost is about \$160 for all available years;
	\item The average number of vehicles insured per contract is around 1.70;
	\item On average, a vehicle is insured for 2.75 contracts.
\end{itemize}

We also split the data into a training set and a test set. To maintain the dependency  between the contracts and the vehicles of the same policy, the training and test set were formed by policy number selection. For example, if a policy is in the training set, it means that all vehicles and contracts in that policy are in the same  training set. Thus, 75\% of the policies were assigned to the training set and 25\% to the test set. These correspond respectively to 75\% and 25\% of all observations. We made these splits by ensuring that we had the same claims frequency  and the same average claims severity in each of the two sets.

\subsubsection{Available Covariates}
\begin{figure}[H]
	\begin{center}
		\includegraphics[scale=0.42]{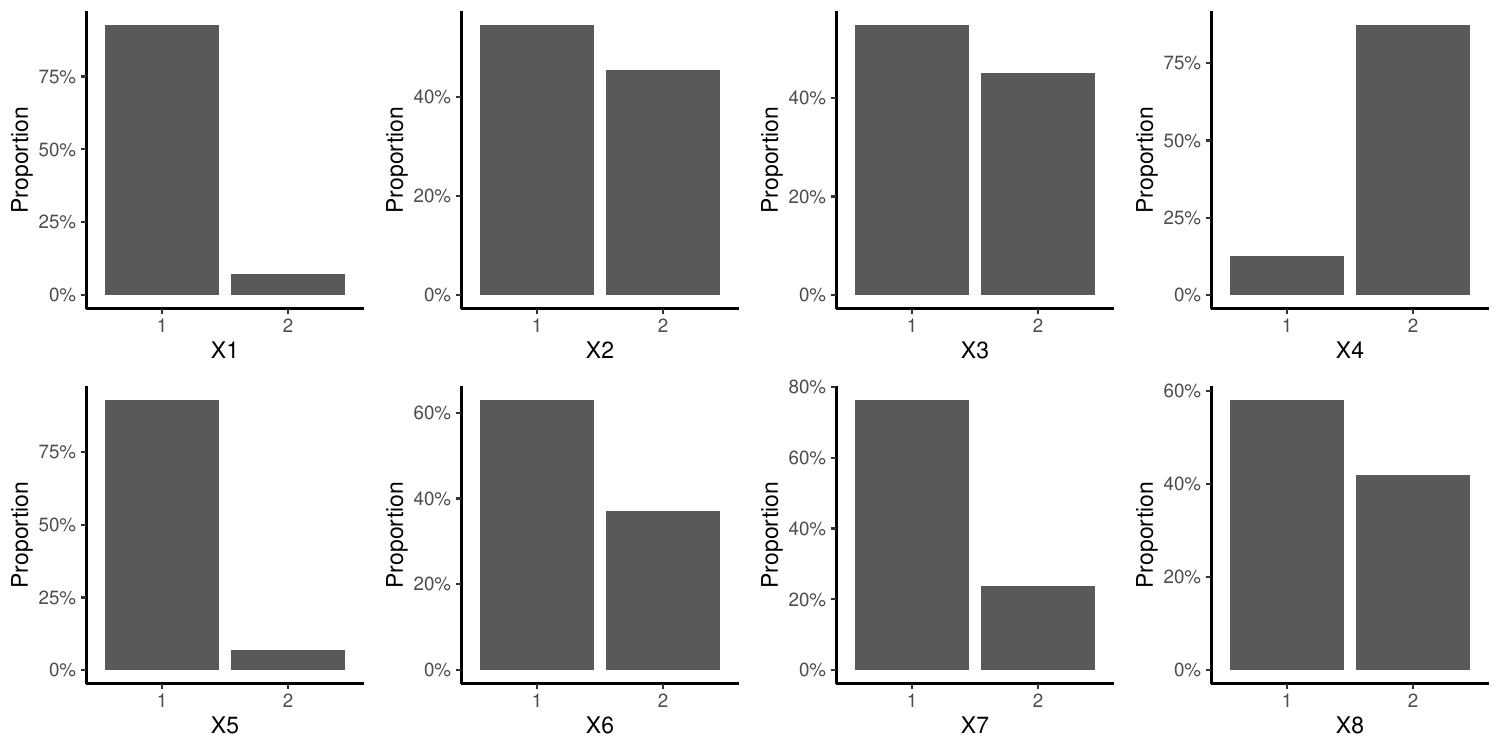}
		\caption{Distribution of covariates}
		\label{covariatesdescription}
	\end{center}
\end{figure}

We have several characteristics  for each vehicle and for each contract. In order to illustrate the impact of segmentation in rating, we select eight of these characteristics as covariates. For confidentiality reasons, but also because this is not the focus of the paper, these covariates are simply labelled as $X1- X8$ and their meaning is not explained. A summary of the proportions of each modality of these variables is given by Figure \ref{covariatesdescription}. To be consistent with the rating approach usually used in practice, which is also often used in the actuarial scientific literature, we have chosen classic risk characteristics related to the sex and age of the insured, the use of the vehicle or the type of vehicle driven, for example. We did not seek to artificially inflate residual heterogeneity from risk characteristics that are not used in pricing.  Thus, we consider that the pricing model developed with the chosen covariates is representative of standard pricing models.

In addition to the selected covariates, indicator variables for each of the calendar years of the contracts were included in the modelling.

\subsubsection{Impact of Past Insurance Experience}
\label{5types}

Although we have 13 years of experience, we use a portion of the database to create a claims history for all insureds. It should be noted that many insureds in the database during the first year have a claims history with the insurer. However, this claims history is not available owing to the structure of the data. Therefore, the first six years of the database are used exclusively to obtain the claims history of each insured, and only the following seven years are used for modelling purposes. For consistency purposes, a fixed window of six years in the past is always used to calculate past claims statistics, $n_{i, \bullet, \bullet}$ and $\kappa_{i, \bullet, \bullet}$, for each of the insureds and each contract. In other words, this means the BMS level of a contract $t$ depends only on the claims experience of contracts $t-1, \ldots, t-6$ and not on the clains experience of the contracts $t-7$, $t-8$ $\ldots$. The impact of this choice of window on the models used is minimal, but it does mean that the Markovian property for a single contract, defined in Equation (\ref{simpletransition}), no longer holds. See \citet{boucher2022} for a study of the window of experience to be used in predictive ratemaking.

A classic quote from \cite{lemaire} is that if only one segmentation variable were to be used for the rating, it should be based on claims experience. For our preliminary analysis of the impact of claims history on premiums, we create six groups of contracts according to their past experience. The first three groups of contracts are based on the number of past contracts ([0.1], [2.3] or [4.5]), and contain only those insureds who could be qualified as inexperienced. We can also call them the new insureds or new policies. The last three groups of policies include only insureds who have been observed for six years or more. The difference between the three groups is based on claims history: the insured in group E and F have filed claims at least once in the past while group D insureds have not filed  claims in the last six years. Table \ref{ExpGroupTable} summarizes the groups of contracts and indicates the frequency, severity and loss cost ratios. These ratios are obtained by dividing the average claims frequency, average claims cost and average loss cost by the corresponding average of each group.

\begin{figure}[H]
	\begin{center}
		\includegraphics[scale=0.42]{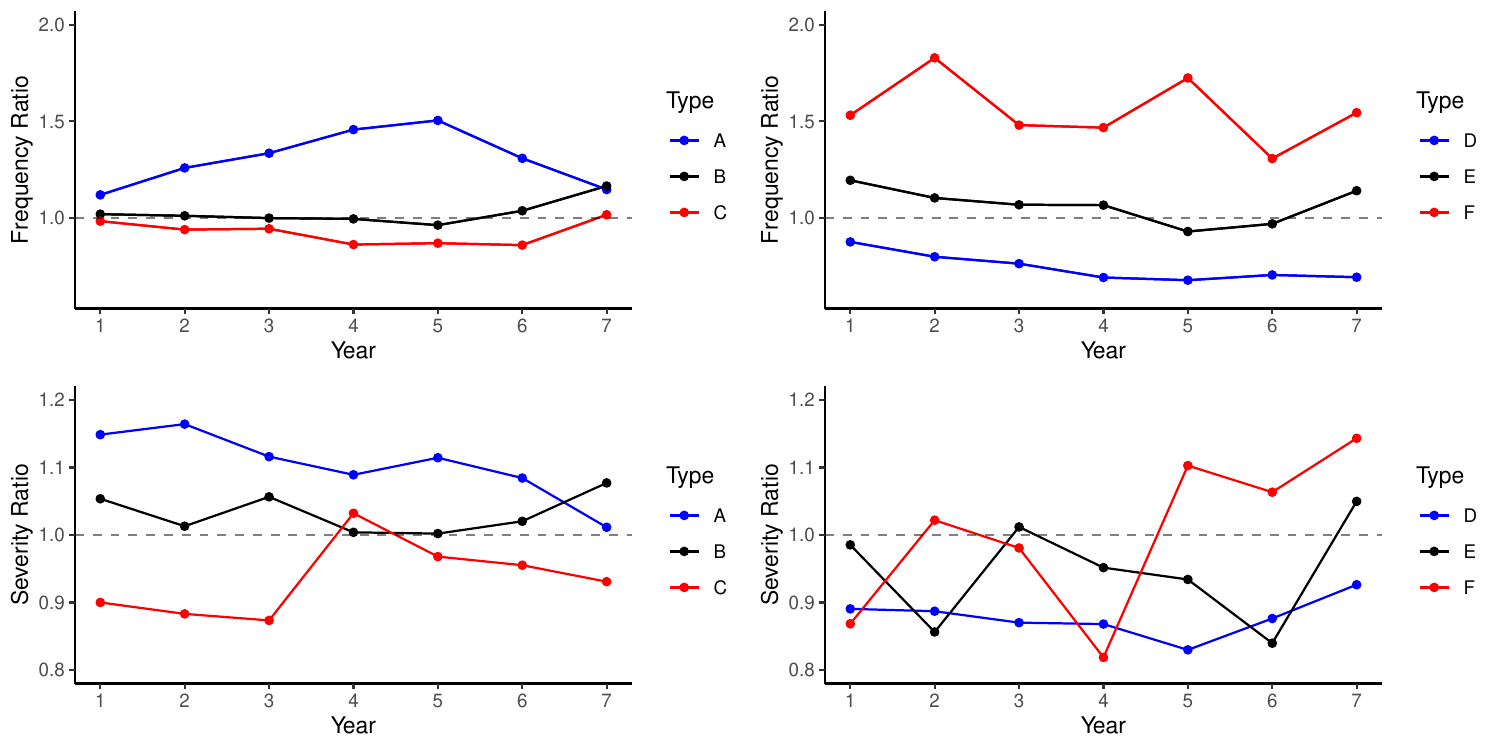}
		\caption{Average claim frequency and severity by group}
		\label{statbygroup}
	\end{center} 
\end{figure}

For each of the seven years studied, Figure \ref{statbygroup} shows the frequency and severity ratios for each group. For a given calendar year or for all years combined, the \textbf{Frequency Ratio} is defined as the ratio of the observed frequency for a group to the observed frequency of the portfolio. This value indicates how much better or worse a group of policyholders is than the portfolio average. The \textbf{Severity Ratio} and the \textbf{Loss Cost Ratio} are defined in the same way.

\begin{table}[ht]
	\centering
	\begin{tabular}{|cc|cc|ccc|}
		\toprule
		& Type  & Past Experience & $n_{i, \bullet, \bullet}$  & Frequency Ratio & Severity Ratio & Lost Cost  Ratio \\
		\midrule
		New insureds &  A& $[0,1]$ & - & 1.303 & 1.110 & 1.446 \\ 
		& B & $[2,3]$ & - & 1.025 & 1.027 & 1.052 \\ 
		& C & $[4,5]$ & - & 0.915 & 0.950 & 0.870 \\ \hline
		Other insureds & D & $\ge 6$ & $0$ & 0.743 & 0.863 & 0.641 \\ 
		& E & $\ge 6$ & $1$ & 1.061 & 0.936 & 0.993 \\ 
		& F & $\ge 6$ & $\ge 2$ & 1.572 & 0.989 & 1.555 \\ 	
		\bottomrule  
	\end{tabular}
	\caption{Group of contracts by past experience}
	\label{ExpGroupTable}
\end{table}

 Although the impact of covariates may need to be considered in order to better understand the statistics shown in Table \ref{ExpGroupTable} and Figure \ref{statbygroup}, it is still relevant to comment directly on each group.

\paragraph{Type A:} We observed that new policyholders in Group A have a much worse claims experience than other groups, in terms of both frequency and severity. With a claims frequency 30.3\% higher than average, and an 11.0\% higher severity, the total burden of Group A policyholders is approximately 45\% higher than the portfolio average. 

\paragraph{Type B:} Group B insureds, with only 1 or 2 years of experience more than the to Group A insureds, seem to differ from the latter. Indeed, the curves illustrated in Figure \ref{statbygroup}, representing their loss experience in frequency and severity compared to the portfolio average, are close to one. The insureds of Group B have a higher claims frequency and claims severity than the insureds of Group C, who have one or two years more experience than Group B.

\paragraph{Type C:} Group C policyholders have four or five  years of past experience. They may or may not have had claims during those years. However, when we look at Figure \ref{statbygroup}, their average claims frequency and severity are better than the averages of the portfolio. We can see, based on the number of years of experience in the company, that a minimum of about four years is necessary to have insureds with claims experience similar to the average of the portfolio.

\paragraph{Type D:} Insureds in this group have insurance experience but have never filed a claim in the last six years. What can be quickly noticed from the figure and the table is that experienced insureds who have not claimed in the last six years (Type D) have a lower claims frequency  than other insureds. Surprisingly, this same group of insured also has a better severity than the others.

\paragraph{Type E:} Group E policyholders are those who have insurance experience but have filed a claim once in the last six years. These insureds have a claim frequency comparable to the new insureds with two and three years of insurance experience. In contrast, their average  claims costs are generally lower than new insureds and the average claims cost of the portfolio.

\paragraph{Type F:} Finally, Group F insureds also have insurance experience but have made at least two claims in the last six years. These insureds are the ones who produce the most interesting claims statistics. In fact, they have a 57\% higher frequency of claims than the portfolio average. They claim  more than the new policies of Group A. However, their average claims cost is lower than the average claims cost of the portfolio compared to the insureds in Group A.  

Through these analyses, we show  how important it is to correctly identify the contracts of new insureds (especially those in Group A) from those of Group D because the insureds of these two groups have the same value for $n_{\bullet}$. The use of a covariate, counting the number of past contracts without claims $\kappa_{\bullet}$ is then justified. 

Finally, to better understand how the past insurance experience impacts each target variable, it is necessary to model their distribution. One can also use other covariates to have flexible rating models.

\subsection{Covariate selection}

The data were used to adjust three types of models for frequency, severity, and loss cost: 

\begin{enumerate}
\item A model that will be called \textbf{standard}, which has no component related to past claims experience;
\item The \textbf{Kappa-N} model (Section \ref{KappaNSection}) using covariates $n_{\bullet}$ and $\kappa_{\bullet}$;  
\item The \textbf{Bonus-Malus Scale} (Section \ref{BMSSection}) limiting the claims Score to $\ell_{min}$ and $\ell_{max}$.
\end{enumerate}
	
For each of the models, we considered the same vector of characteristics except for the Kappa-N and BMS models which use also use the covariates  $\kappa_{i,\bullet, \bullet}$ and $n_{i,\bullet, \bullet}$. However, not all risk characteristics consistently have the same impact in our three variables of interest. For example, the frequency of claims is greatly impacted by the characteristics of the insured, such as age and gender, while the severity of collision coverage will usually be more influenced by the characteristics of the vehicle, mainly the value of the vehicle. Therefore, a statistical procedure for selecting covariates seems necessary.

We adopted the \textit{Elastic-net} regularization to select the covariates. This method is seen as a combination of Lasso and Ridge regressions. See \citet{hastie2009elements} and \citet{statilearn} for more details about this approach. One of the advantages if this approach is that it solves the redundancy of variables and the multicollinearity of risk factors. The idea of the procedure is to impose constraints on the coefficients of the model. Excluding the intercept from the procedure, the constraint to be added to the log-likelihood score to be maximized is expressed as:

$$\left(\alpha \sum_{j = 1}^{q + 1} \mid \bm{\beta}^{(.)}_j \mid  +\frac{ (1 - \alpha)}{2}\sum_{j = 1}^{q + 1} \bm{\beta}^{(.)2}_j\right) \le \lambda,~~ \lambda >0,~~0 \leq \alpha \leq 1.$$

This penalty constraint depends on the values chosen for the parameters  $\alpha$ and $\lambda$. If $\alpha = 0$, the \textit{Elastic-net} is equivalent to a Lasso regression. In contrast, if $\alpha = 1$, it is equivalent to a Ridge regression. For each studied model, the optimal values of  $\lambda$ and $\alpha$ were obtained by a cross-validation using deviance as a selection criterion.

\subsection{Ftting Statistics and Prediction Scores}

Table \ref{modelscomparisonscpg} shows the fit results of the models based on the training set, and the prediction quality based on the test set. The number of parameters used in the model (after applying the
procedure by elastic net), the log-likelihood, as well as the AIC (Akaike information criterion) and BIC (Bayesian information criterion) for each of the models are indicated. To evaluate the prediction quality based on the test set we avoided using least squares because they are not always adequate for frequency, severity, or loss cost statistics. A logarithmic score $SL$ representing the negative loglikelihood value on the test set with the parameters estimated on the training set is used instead. More formally, if we denote by $\bm{\hat{P}}$ the estimated parameters for each of the models, the logarithmic prediction score is calculated as: $SL = - \log(f(\bm{\hat{P}}|Test~ set))$, where $f()$ is the probability density of function of each target variable. As with least squares, the aim is to obtain the smallest value of $SL$ on the test set in order to control the over-fitting resulting from the estimation of the parameters on the training set. Optimal use of this score implies the estimation of all parameters by maximizing the likelihood function and not only the parameters associated with the mean of each target variable.

\begin{table}[H] 
	\centering
	\begin{tabular}{ *{16}{c} } 
		\toprule 
		\multicolumn{2}{c}{} & &\multicolumn{3}{c}{Train} & Test\\
		\multicolumn{2}{c}{Model}&Number of parameters&Log-likelihood&AIC&BIC& Score\\
		\midrule
		\multirow{4}{*}{Standard}
		&\textit{Poisson}&\textit{15}&\textit{-97,786}&\textit{195,602
		}&\textit{195,782}&\textit{32,939}\\
		&\textit{Gamma}&\textit{15}&\textit{-187,229}&\textit{374,489
		}&\textit{374,606}&\textit{63,396}\\  \cline{2-7}
		&CPG&30&-284,252&568,564&568,924&95,982\\
		&Tweedie&31&-281,849&563,760&564,131&95,221\\
		\midrule
		\multirow{4}{*}{Kappa-N}
		&\textit{Poisson}&\textit{17}&\textit{-97,426}&\textit{194,887
		}&\textit{195,090}&\textit{32,811}\\
		&\textit{Gamma}&\textit{16}&\textit{-187,189}&\textit{374,410}&\textit{374,536}&\textit{63,384}\\  \cline{2-7}
		&CPG&33&-283,885&567,836&568,232&95,851\\
		&Tweedie&33&-281,490&563,046&563,441&95,092\\
		\midrule
		\multirow{4}{*}{BMS}
		&\textit{Poisson}&\textit{19}&\textit{-97,421}&\textit{194,876}& \textit{195,079}& \textit{32,812}\\
		&\textit{Gamma}&\textit{18}&\textit{-187,189}& \textit{374,413}& \textit{374,555}&\textit{63,382}\\  \cline{2-7}
		&CPG&37&-283,881&567,836&568,280&95,852\\
		&Tweedie&34&-281,487&563,042&563,449&95,097 \\
		&Tweedie's CP&37&-282,151&564,376&564,819&95,260\\
		\bottomrule 
	\end{tabular} 
	\caption{Model comparisons }
	\label{modelscomparisonscpg}
\end{table}

\subsection{Comparison between Models}

\subsubsection{CPG vs Tweedie}

The first angle of analysis is to compare the fit and prediction quality of the CPG model with the Tweedie model. Some analyses of the insurance data showed that the gamma distribution was not rejected by a hypothesis test (the QQ-plots are available in Appendix B \ref{apendixA}), but the Poisson distribution for the number of claims is not ideal. Given the convergence of the  Poisson model estimators, however, the assumption of a Poisson distribution for the annual claims number will be retained, which will allow us to continue the analysis with Tweedie.

As mentioned by \cite{delong2020}, to use likelihood based criteria to compare the two approaches (CPG and Tweedie), the data samples must be the same in each model. So, using our remark from \ref{remarques}, we get the adjustment statistics and the prediction scores of the two models as given by Table \ref{modelscomparisonscpg}. 

\citet{delong2020} also argued that the fit quality of the Tweedie is always better than that of the CPG if the same covariates are used in both models and under other assumptions that we have considered here as well. However, we do not have the same covariates in the two models due to the Elastic-net procedure and the addition of the claims score or BMS level in the mean parameter modelling in each model. Considering the adjustment statistics and the prediction scores in Table  \ref{modelscomparisonscpg}, the Tweedie models are  better than the CPG models.  

For only the BMS case, we consider a Tweedie model (\textbf{Tweedie' s CP}) with the same covariates selected as in the CPG model in line with \citet{delong2020}. This Tweedie model is better than CPG, as \citet{delong2020} assert, but our proposed Tweedie model is still better than Tweedie's CP model. 

\subsubsection{Standard  vs Kappa-N and BMS}

Another angle of analysis, more specific to what we have presented in the paper, is related to the modelling of past experience. First, the analysis should be divided according to the chosen distribution, and then a summary analysis should be done of all the models combined.

\paragraph{Poisson (Frequency):} The introduction of a claim score or BMS level into the mean parameter of a Poisson distribution is not new. Nevertheless, it is still interesting to see that the Kappa-N and BMS models significantly improve the AIC and BIC statistics, compared to the standard models. The prediction score is also improved if one switches from the standard model to the Kappa-N or BMS model. The differences in the fit statistics and prediction score of the Kappa-N model and the BMS model are minimal. Knowing that the Kappa-N model is difficult to use in practice, it is interesting to see that the cost of having a predictive pricing model that has a potential for use is negligible.

\paragraph{Gamma (Severity):} Modelling severity based on the number of past claims is not a very common approach in actuarial science. As we saw earlier in the severity analysis based on five groups of insureds, severity approaches that included a loss experience component were expected to perform well. This is what we are seeing: the Kappa-N and BMS models produced better values of AIC, BIC, and a better logarithmic prediction score than did the standard model. Considering that the standard model does not use claims history in the premium calculation, this result is very interesting
since it seems to indicate that there is value for insurers in including a component of discount and surcharge on severity based on past claims. As with frequency, the observed differences between the values of AIC, BIC and logarithmic $SL$ score are very small between the BMS approach and the Kappa-N approach, showing once again that the requirement to have a practical approach is not very restrictive.

\paragraph{CPG (Loss cost):} The CPG model is the combination of the frequency and severity approach discussed above. Both the frequency and severity approaches favor the Kappa-N and BMS models. Thus, it is obvious that both of these approaches will be better than the standard approaches. We also notice that the results are much better with the BMS model except for the BIC criterion, which causes a greater number of parameters to be penalized, at 37 compared to 33 for Kappa-N.

\paragraph{Tweedie (Loss cost):} As with severity modelling, the use of the number of past claims to model the loss cost is not a very common approach in actuarial science and it is therefore very interesting to check whether this generalization of the approach is relevant. Analysis of the AIC, BIC and $SL$ tends to show that the addition of elements related to past insurance experience helps to better segment the risk for collision coverage. Indeed, the values obtained for the Kappa-N and BMS approaches strongly favor these models, compared to the standard approach. Just like before, the BMS model performed better than the Kappa-N except for the BIC criteria and the prediction score, where a difference was observed for the Kappa-N model.

\subsection{Analysis of Premiums} 

\subsubsection{Estimated Parameters } 

Table \ref{parbeta1} in Appendix A \ref{apendixA} shows the estimated values of the $\beta$ used with the selected covariates for the Standard model and the BMS model. For frequency, severity, and loss cost, there is a marked difference between some estimators. The impact of adding components linked to a BMS level on parameters related to segmentation variables had already been observed and analyzed by \citet{boucher2014} for frequency analysis. We will not elaborate further, especially since the same explanations of the reasons for these differences apply to the analysis of severity and loss cost.

Above all, it is necessary to analyze in a little more detail the values of the estimators of the parameters related to the claim experience. Table \ref{OtherParms} shows the value of the parameters related to past claims experience for the Kappa-N model and the value of the structural parameters for the BMS approach. For frequency, severity, and loss cost, the table shows that the impact of estimating the parameters $\gamma_0$ and $\Psi$  is small when minimum and maximum BMS limits are added, i.e. $ell_{min}$ and $\ell_{max}$.

\begin{table}[ht]
	\centering
	\begin{tabular}{|c|ccc|ccc|}
		\toprule
		& \multicolumn{3}{c|}{Kappa - N} & \multicolumn{3}{c|}{BMS}   \\
		Parameters          & Poisson &        Gamma        & Tweedie   & Poisson &     Gamma     & Tweedie  \\
		\midrule
		$\gamma_0$& 0.081& 0.025& 0.107& 0.094& 0.026& 0.112\\
		$\psi$    & 2.899& 2.073& 2.728&      3&    2&     3\\
		$[\ell_{min}, \ell_{max}]$&-&-&-&[95,106]&[94,100]&[95,104]\\
		\bottomrule 
	\end{tabular}
	\caption{Estimation of the other parameters of the Kappa-N and BMS models}
	\label{OtherParms}
\end{table}

\begin{figure}[H]
	\begin{center}
		\includegraphics[scale=0.42]{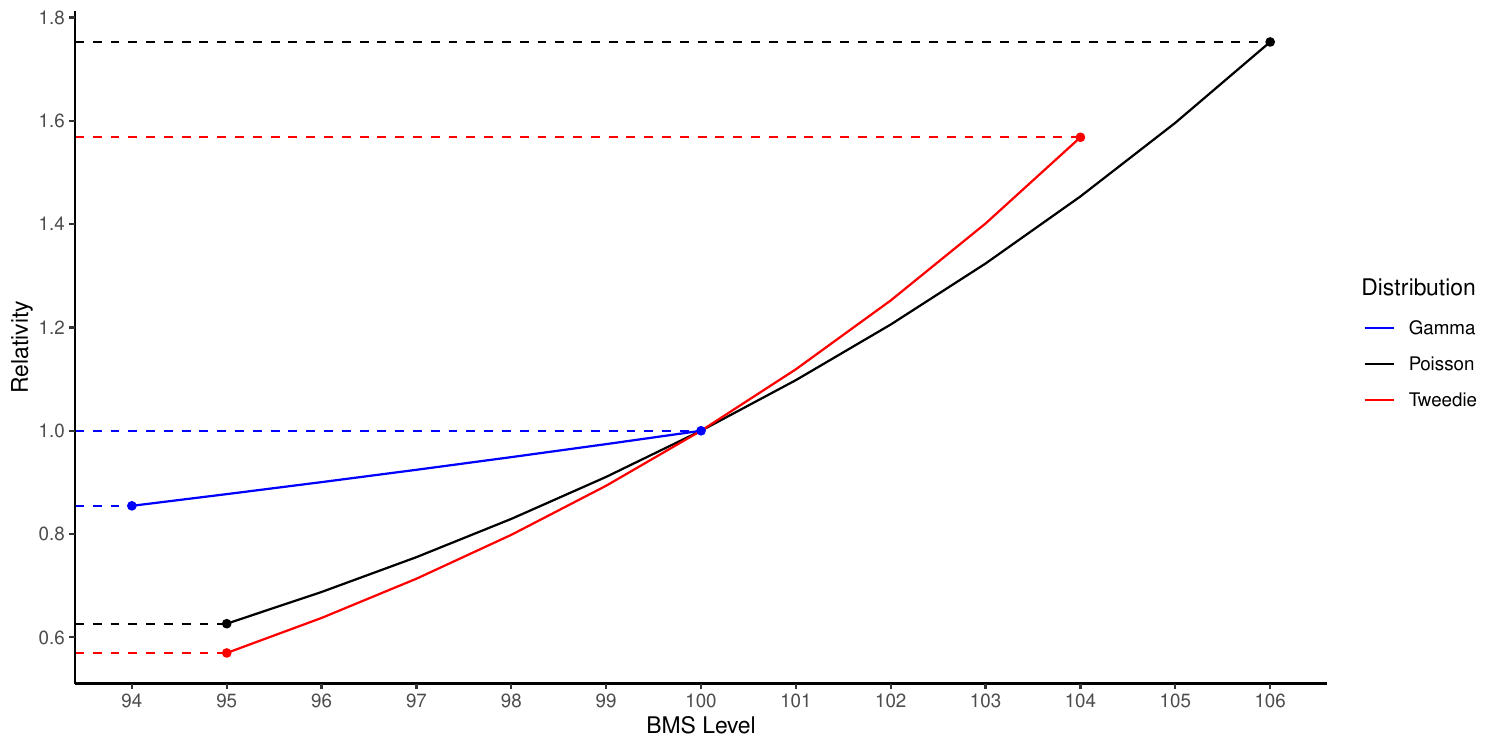}
		\caption{BMS relativites}
		\label{BMSRel}
	\end{center}
\end{figure}

The results in Table \ref{OtherParms} indicate that the jump parameter for a past claim is the same for the frequency and loss cost ($\Psi^{N} = \Psi^{Y} = 3$), but this parameter is different for severity ($\Psi^{Z} = 2$). The relativity parameter  $\gamma_0$ of the frequency is also closer to that of the loss cost than to that of the frequency. Finally, the frequency model proposes BMS level limits that are slightly larger than the severity or loss cost. To better understand how the experience of past claims impacts policyholders' premiums according to the studied models, we can refer to Figure \ref{BMSRel} which illustrates the relativity curve of the frequency (Poisson), the
severity (gamma) and loss cost (Tweedie) as a function of BMS level. The impact of the parameters $\Psi$, $\gamma_0$, $\ell_{min}$ and $\ell_{max}$ can all be observed simultaneously in the same figure:

\begin{itemize}
	\item It shows that the range of possible penalties for severity (in blue) is much smaller than those for frequency (red) and loss cost (black).
	
	\item The maximum penalty for frequency is higher than the maximum penalty for loss cost, which is much higher than the penalty for severity. We reach the same conclusion by comparing the maximum discount obtained in each model. 
	
\end{itemize} 

In the CPG model, an insured's total premium is the frequency multiplied by the severity. For the premium calculation of this model, the BMS levels of frequency and severity must be calculated. Thus, it is not possible to illustrate the BMS relativities of the CPG model in two dimensions, as is done in Figure \ref{BMSRel}. To compare BMS surcharges, discounts, and minimum and maximum relativities, BMS models of frequency and severity should be combined. The result of this comparison is shown in Table \ref{CombinRel}. The direct comparison
between the CPG model and Tweedie model shows that the surchages of the two approaches for a claim are similar: a premium increase of 40.1\% compared to an increase of 39.5\%. The discounts for a claims-free year are also similar: -11.3\% versus -10.6\%. The most striking difference between the two models is revealed above all at the level of relativity of each model. In the CPG approach, an insured with a lot of past claims might have a surchage of more than 175.3\% while this surchage is limited to 156.8\% for the Tweedie model. In contrast, the maximum discount is comparable in both models.

\begin{table}[ht]
	\centering
	\begin{tabular}{|c|ccc|c|}
		\hline
		& Poisson & Gamma & CPG & Tweedie \\ 
		\hline 
		Surcharge per Claim & 0.324 & 0.054 & 0.395 & 0.401 \\ 
		Claims Free Discount & -0.089 & -0.026 & -0.113 & -0.106 \\ 
		Min. \& Max. Relativities  & [0.626, 1.753] & [0.855, 1.000] & [0.535, 1.753] & [0.570, 1.568] \\ 
		\hline
	\end{tabular}
	\caption{Impacts of past claims for all BMS Models}
\label{CombinRel}
\end{table}

\subsubsection{Numerical Example}

To better illustrate the similarities and differences between the CPG model and the Tweedie model, we will use the estimated parameters of the models and thus assume four insureds with the claim history shown in Table \ref{table:3insureds}. Each insured was observed for twelve consecutive years. The first insured has not filed a claim during the 12-year periods, insured \#2 is a bad driver who claims frequently, insured \#3 filed many claims in the first three years, but the number diminished during the other years, and the last insured has a deteriorating driving experience. It will be assumed that all insured persons start at the BMS level of 100 at year 0, for the bonus-malus scale of frequency, severity and loss cost. As was done with the insurance data used earlier, a 6-year window is assumed for the calculation of levels and the first 6 years are used to create a claims history. We will analyze the resulting premiums for each insured in years 7 to 12.

\begin{table}
	\begin{center}
		\begin{tabular}{|c|c|c|c|c|c|c||c|c|c|c|c|c|}\hline
			Insured & \multicolumn{12}{c|}{Years ($t$) } \\
			$i$ & 1 & 2 & 3 & 4 & 5 & 6 & 7 & 8 & 9 & 10 & 11& 12\\ \hline
			1 & 0 & 0 & 0 & 0 & 0 & 0 & 0 & 0 & 0 & 0 & 0 & 0 \\ 
			2 & 2 & 0 & 1 & 2 & 0 & 1 & 2 & 0 & 1 & 2 & 0 & 1 \\ 
			3 & 4 & 1 & 3 & 0 & 1 & 0 & 0 & 0 & 0 & 0 & 2 & 0 \\ 
			4 & 0 & 2 & 0 & 0 & 0 & 0 & 0 & 1 & 0 & 3 & 1 & 4 \\ \hline
		\end{tabular}
		\caption{Insureds with claims experience}
		\label{table:3insureds}
	\end{center}
\end{table}

\begin{figure}
	\begin{center}
		\includegraphics[scale=0.42]{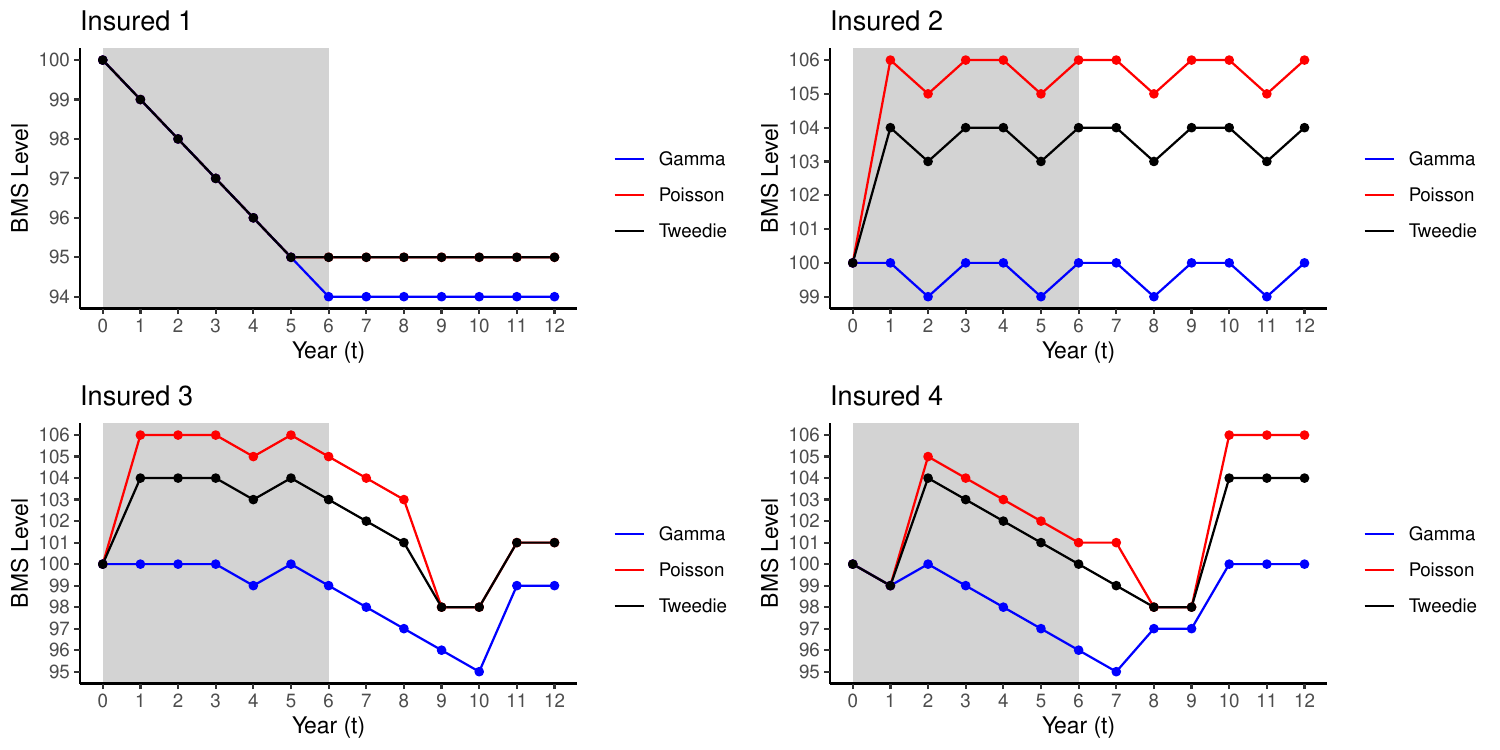}
		\caption{BMS levels for all four fictitious  insureds}
		\label{BMSRelEx}
	\end{center}
\end{figure}

At the beginning of each year $t$, Figure \ref{BMSRelEx} shows the evolution of BMS levels of frequency, severity and loss cost for each of the four fictitious insureds. The grey area of each graph corresponds to the first six years used for the calculation of the initial BMS level.

\begin{figure}
	\begin{center}
		\includegraphics[scale=0.42]{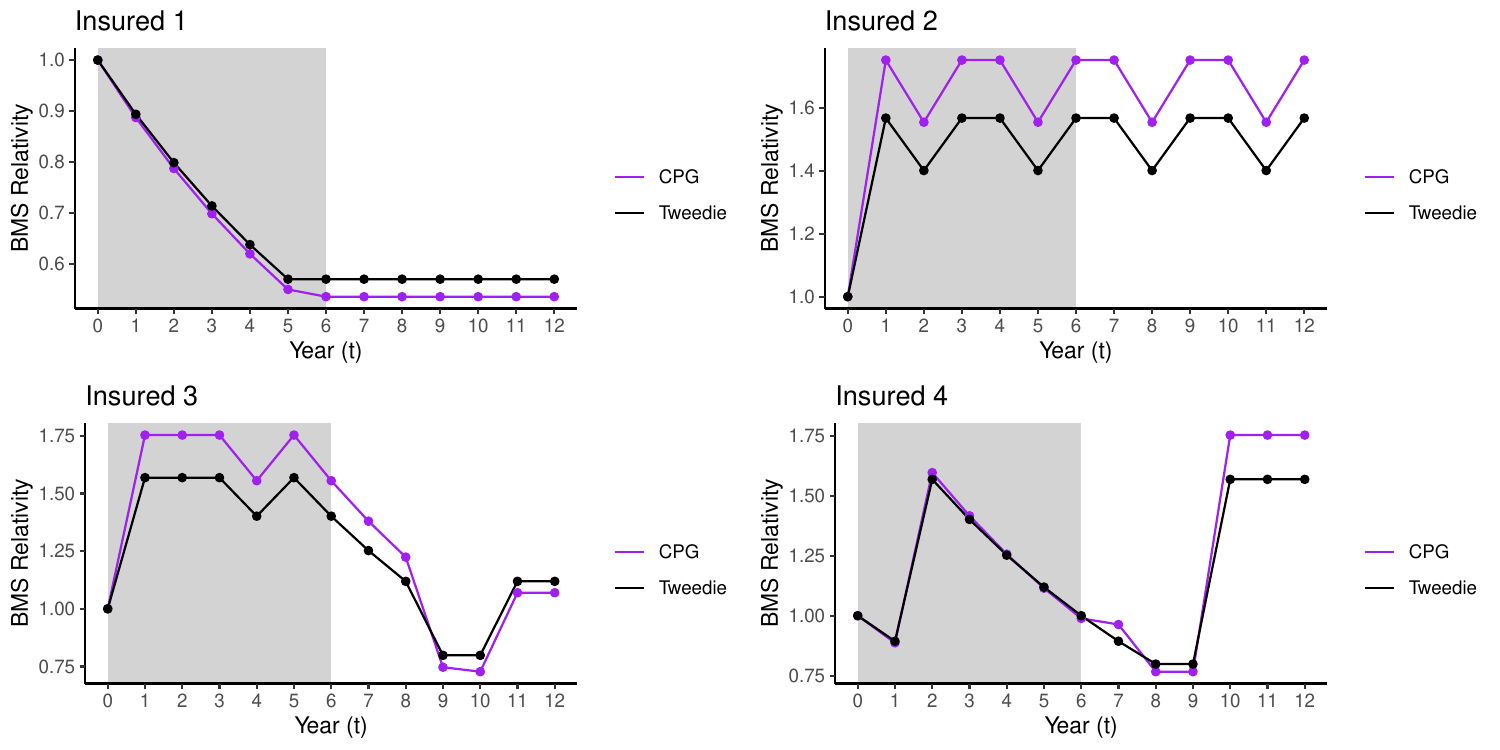}
		\caption{BMS relativities for all four fictitious insureds}
		\label{BMSPrimeEx}
	\end{center}
\end{figure}

Figure \ref{BMSPrimeEx} shows the resulting BMS relativity, where the BMS relativity of the CPG model is the combined effect of frequency and severity. For all years after the sixth contract ($t \ge 7$), we can see that the BMS relativities for each of the two models are similar for the four insureds in the example.  Thus, even though the BMS levels sometimes appear to be different, the combination of severity and frequency means that the relativity obtained is close to that of Tweedie. Of course, there are some differences between the two curves, but the general trend is always the same.

Obviously, the four insureds in the example are fictitious situations, and the supposed history may not have taken place in the database used. The main purpose of the example is to show that despite the imposition of a rigid predictive rating structure, the models are still comparable.

\subsubsection{CPG vs Tweedie}

\begin{figure}
	\begin{center}
		\includegraphics[scale=0.40]{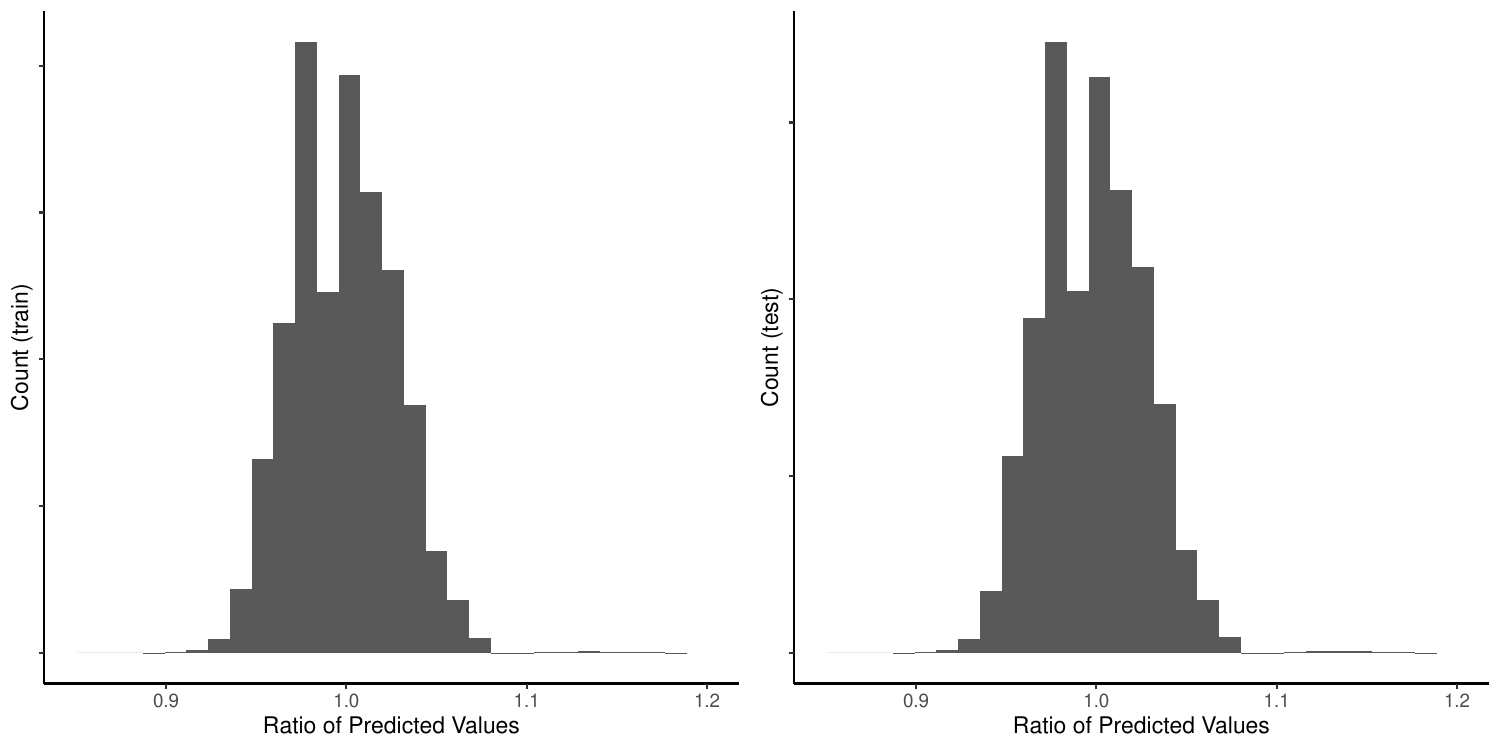}
		\caption{Premium ratio (left: training set, right: test set)}
		\label{Comp2Tweedie}
	\end{center}
\end{figure}

Figure \ref{Comp2Tweedie} shows the ratio between the CPG premium and Tweedie for the training set and test set. The distribution is similar for both parts of the dataset. We can also see that the premium ratio is around 1 but that spreads of minus 95\% or more than 110\% exist in the portfolio. The choice of a type of model for predictive pricing thus has a significant potential impact.

\subsection{Predicted and Observed Loss Cost}

\subsubsection{Types of Insureds}

\begin{figure}
	\begin{center}
		\includegraphics[scale=0.42]{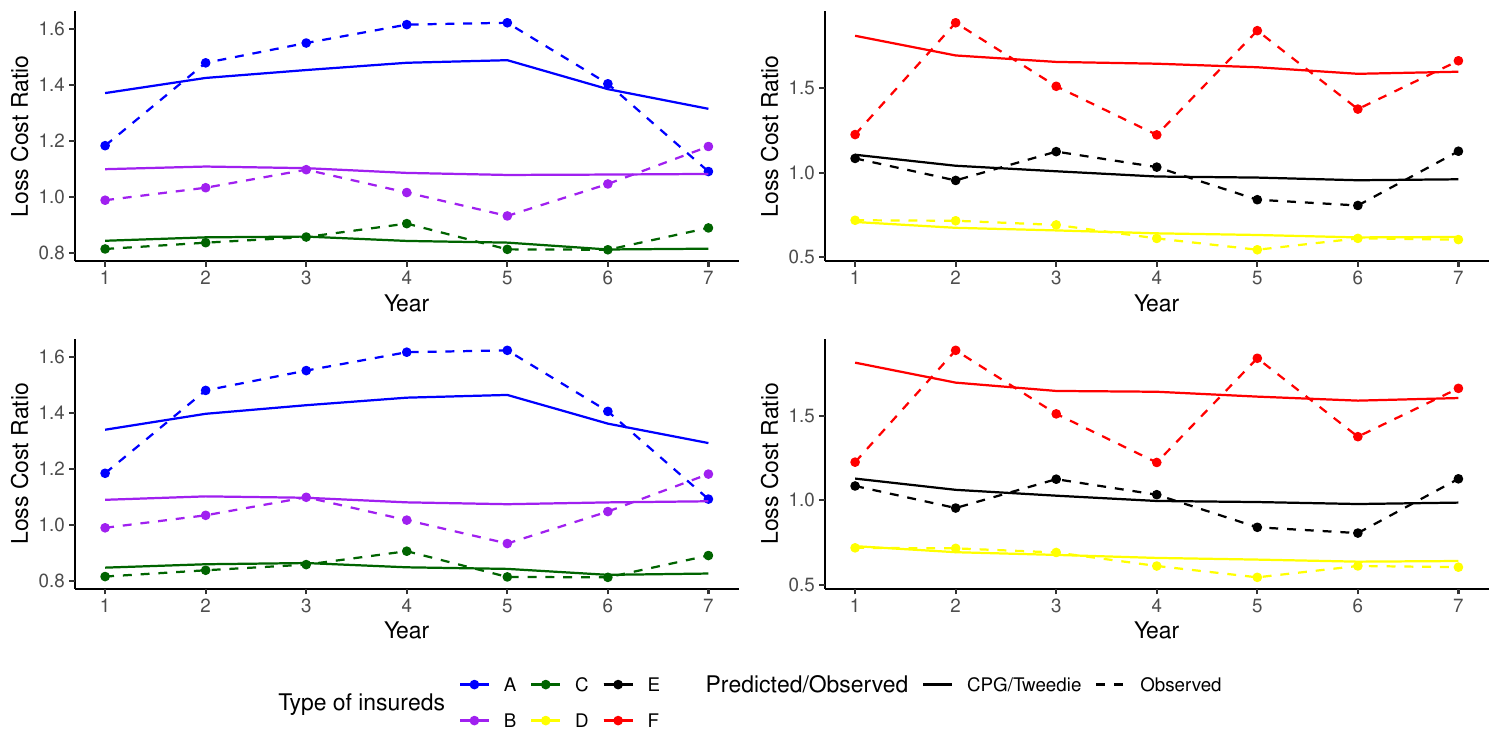}
		\caption{Loss Cost for the test set (top: CPG, bottom: Tweedie, left: Type A-B-C, right: Type D-E-F )}
		\label{5typesFig}
	\end{center}
\end{figure}

A relevant way to compare BMS models with the two underlying distributions (CPG and Tweedie) is to check the fit between what is observed and what has been predicted for each of the models according to the type of contracts. By taking the five types of policyholders defined in Section \ref{5types}, we can potentially see the type of contract for which the two Tweedie models could be improved. Table \ref{LC5types} below shows the ratio of the predicted loss cost to the annual average for the five types of policyholders. Figure \ref{5typesFig}, in contrast, illustrates the observed pure charge ratio and those predicted for both models: the two graphs at the top are for CPG and the ones at the bottom are for Tweedie. For the test portion of the database, the graphs on the left refer to type A, B and C insureds (and therefore what could be called new insureds), and those on the right indicate the result for type D and E insureds, i.e. insured with at least six years of insurance experience.

This analysis by type of insured makes it possible to see the type of insured that seems to be best or worst predicted by the different models. The two approaches, CPG and Tweedie, predict the loss costs of policyholders whose average total charge is close to or smaller than that of the portfolio. This prediction is almost perfect for Group D. In contrast, for policyholders who have an average loss cost higher than that of the portfolio, the costs are generally underestimated by both approaches.

\begin{table}[ht]
	\centering
	\begin{tabular}{c|ccc|ccc}
		\hline
		& \multicolumn{3}{c}{Training Set} & \multicolumn{3}{c}{Test Set} \\
		Type & Observed & CPG & Tweedie  & Observed & CPG & Tweedie\\ \hline
		A & 1.482 & 1.407 & 1.396 & 1.434 & 1.408 & 1.399 \\ 
		B & 1.068 & 1.108 & 1.105 & 1.047 & 1.102 & 1.099 \\ 
		C & 0.896 & 0.854 & 0.856 & 0.861 & 0.851 & 0.853 \\ 
		D & 0.644 & 0.650 & 0.670 & 0.641 & 0.649 & 0.668 \\
		E & 0.933 & 1.012 & 1.034 & 1.013 & 0.999 & 1.020 \\ 
		F & 1.393 & 1.658 & 1.657 & 1.613 & 1.669 & 1.668 \\ 
		\hline
\end{tabular}
\caption{Loss Cost Ratio for all types of insureds}
\label{LC5types}
\end{table}

\subsubsection{BMS Levels}

\begin{figure}
	\begin{center}
		\includegraphics[scale=0.39]{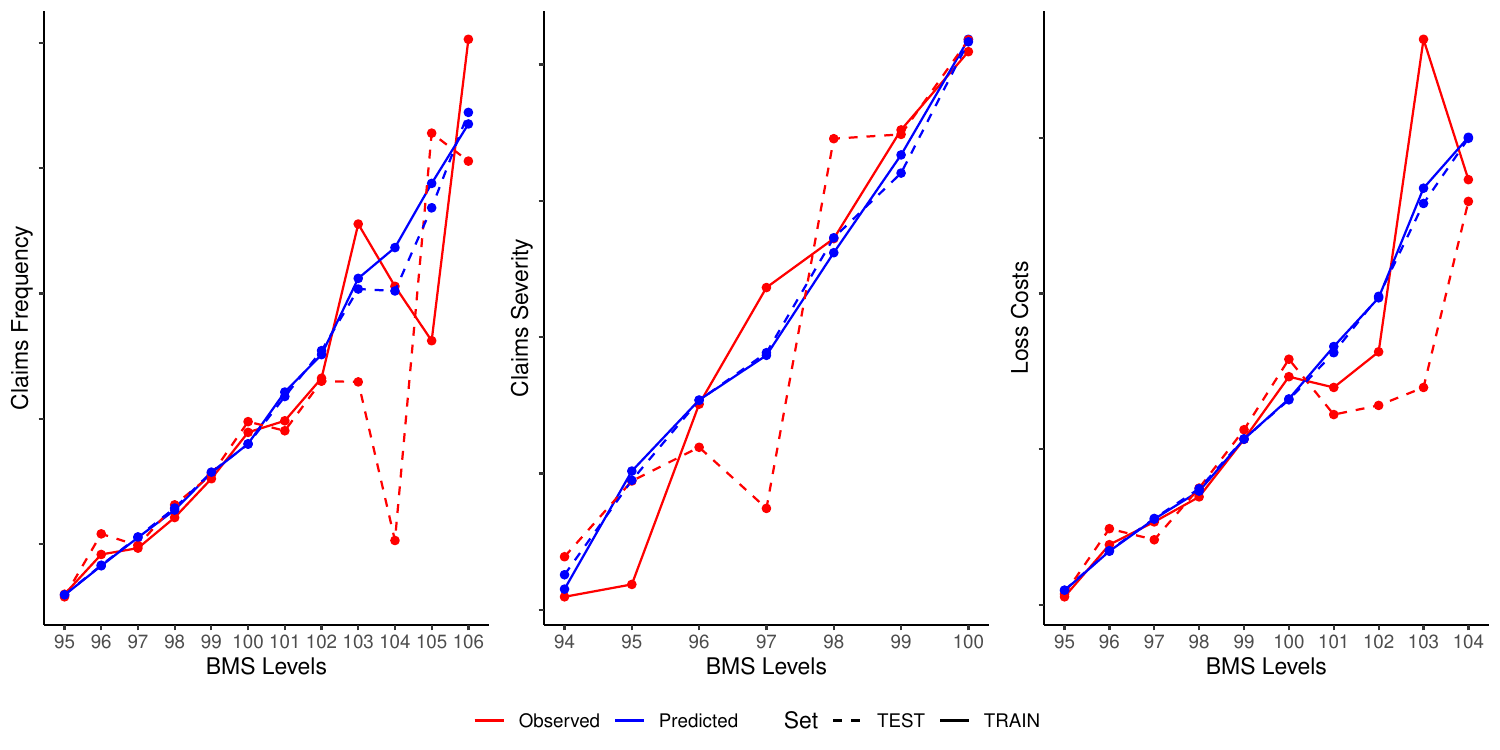}
		\caption{Predicted vs Observed for the claims frequency (left) and the claims severity (right)}
		\label{PredObsLevels}
	\end{center}
\end{figure}

It may also be interesting to check the fit between the observed and the predicted values according to the BMS level of the contract. Figure \ref{PredObsLevels} illustrates this fit for frequency, severity, and loss cost. The blue curves represent the predicted mean relativity while the red curve represents the observed relativities. The solid lines are for the training set while the dotted lines represent the results of the test set. For the frequency of claims, we see that the difference between predicted and observed relativities is minimal for contracts with BMS levels below 103. For levels 103 and above, where there are far fewer insureds, we can see that the general trend of the model is in line with the average of what has been observed. For the severity model, the relativity curve clearly shows a decrease when the BMS level decreases, which is what the pricing model also assumes. The difference between the observed and the predicted values is more variable for severity than for frequency. Finally, the gap between the observed relativities and the relativities obtained by the Tweedie model for the total charge is close to what was observed for the frequency model: the difference is minimal for lower BMS levels and more variable for higher levels.

\section{Conclusion}\label{conclusion}	

We generalized the paper of \citet{delong2020} by including a predictive ratemaking component in the premium. Our approach can also be seen as an extension of the work of \cite{boucher2022} by considering the severity and the loss cost as target variables in addition to the frequency. In other words, our objective was to compare the BMS models to the standard models when the CPG and Tweedie are used as underlying distributions. Our main conclusions and remarks are as follows. 

First, in the BMS model with CPG as an underlying distribution, we found that BMS level impacts the frequency and severity components of the CPG differently. Although both are positively related to the BMS level, the impact is stronger for the frequency component than the severity component. This results in the surcharge and discount being significant in the frequency component than  severity. Finally, by comparing the relativities, the frequency component penalizes insureds who have made a lot of claims in the past more than the severity component does.

Second, in the BMS model with Tweedie as an underlying distribution, we found positive dependency between the BMS level and the corresponding premium. We also noted that the surcharge by claim and the claims-free discount are comparable in both the Tweedie and CPG cases. However, the CPG model penalized insureds who have made a lot of claims in the past more than the Tweedie model did.  

Finally, we obtained the same conclusions as \cite{boucher2022} about the BMS model. All BMS models considered in our analysis have better quality data adjustment and data prediction than the standard approaches do. These statistics are better when the Tweedie is used as an underlying distribution compared with the CPG. In addition, the Tweedie model (\textbf{Tweedie}) with a unique BMS level is better than the Tweedie model (\textbf{Tweedie' s CP}), which uses the BMS levels  obtained in the CPG model. 

We conclude the paper with some remarks about the underlying distributions used. First, there are other excellent distributions for the frequency and the severity modelling. For example, the  Negative-Binomial is preferable to the Poisson. Yet to compare the CPG and Tweedie models, the frequency and severity components of the CPG must be modelled by the Poisson and gamma distributions. Finally, due to the positive probability density function of loss cost when this loss cost is zero, a mixture distribution (discrete and continuous) may be an adequate choice for the loss cost modellling. This is why we make an appropriate choice of the variance parameter of the Tweedie distribution to allow loss cost modelling.

\newpage
\section*{Appendix A: Estimated Coefficients of the  Mean Parameter}\label{apendixA}

\begin{table}[ht]
	\centering
	\begin{tabular}{|c|c c|c c|c c |c c|}
		\toprule
		          & \multicolumn{2}{c|}{Standard} & \multicolumn{2}{c|}{BMS} & \multicolumn{2}{c|}{Standard} & \multicolumn{2}{c|}{BMS}\\
Parameter  & Poisson & Gamma& Poisson& Gamma &CPG & Tweedie& CPG & Tweedie\\
				          \midrule
$\beta_0$    &-4.409& 8.770&-13.541& 6.227&  4.361&  4.330& -7.314& -6.626\\
$\beta_1$    & 0.135& 0.067&  0.121& 0.062&  0.202&  0.207&  0.183&  0.189\\
$\beta_2$    & 0.031&-0.078&  0.044&-0.068 & -0.047& -0.047& -0.024& -0.032\\
$\beta_3$    &-0.154& 0.000& -0.144&   -  & -0.154& -0.157& -0.144& -0.145\\
$\beta_4$    & 0.192&   -  &  0.220& 0.005&  0.192&  0.187&  0.224&  0.225\\
$\beta_5$    & 0.474& 0.109&  0.405& 0.086&  0.583&  0.587&  0.491&  0.505\\
$\beta_6$    & 0.548& 0.186&  0.528& 0.184&  0.735&  0.735&  0.711&  0.716\\
$\beta_7$    & 0.231& 0.098&  0.188& 0.085&  0.329&  0.342&  0.273&  0.283\\
$\beta_8$    &-0.128&-0.150& -0.054&-0.113&  0.199&  0.238&  0.160&  0.176\\
$\beta_9$    & 0.181& 0.018&  0.160&   -  &  0.199&  0.238&  0.160&  0.176\\
$\beta_{10}$ & 0.343& 0.104&  0.316& 0.084&  0.447&  0.486&  0.400&  0.417\\
$\beta_{11}$ & 0.378& 0.160&  0.347& 0.136&  0.538&  0.576&  0.482&  0.499 \\
$\beta_{12}$ & 0.223& 0.190&  0.191& 0.165&  0.413&  0.450&  0.356&  0.371\\
$\beta_{13}$ &-0.095& 0.190& -0.151& 0.160&  0.095&  0.130&  0.009&  0.189\\
$\beta_{14}$ & 0.106& 0.150&  0.042& 0.126&  0.256&  0.298&  0.168&    -  \\
$\beta_{15}$ &   -  &   -  &  0.094& 0.026&   -   &    -  &  0.094&  0.112\\
$\beta_{16}$ &&&&&   -   &    -  &  0.026&     -\\
		\bottomrule
	\end{tabular}
	\caption{Estimated parameters}
	\label{parbeta1}
\end{table}

\section*{Appendix B: Residual Analysis}\label{apendixB}

\begin{figure}[H]
	\begin{center}
		\includegraphics[scale=0.40]{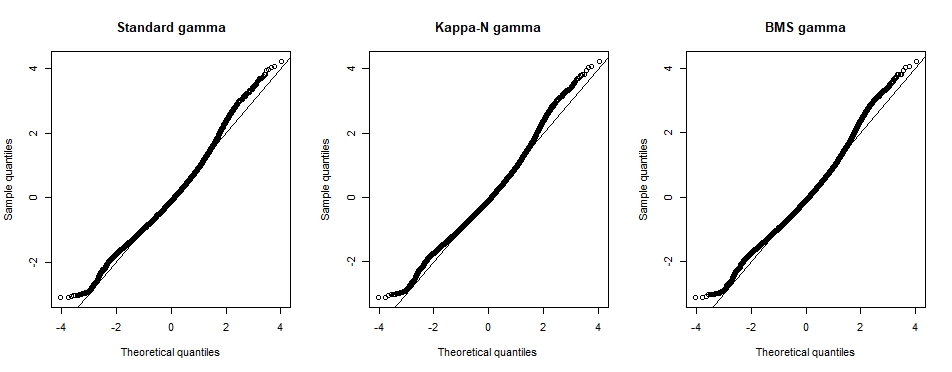}	
		
		\includegraphics[scale=0.40]{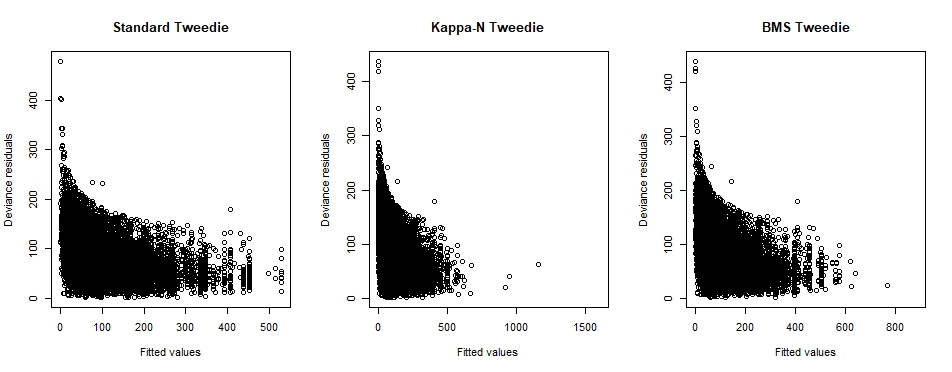}	
		\caption{Cox-Snell residuals for severity and Anscombe residuals for loss cost} 
		\label{gammaadequacy}
	\end{center}
\end{figure}

\bibliography{bibtex}

\begin{thebibliography}{}

\bibitem[\protect\astroncite{Berm{\'u}dez et~al.}{2018}]{bermudez2018allowing}
Berm{\'u}dez, L., Guill{\'e}n, M., and Karlis, D. (2018).
\newblock Allowing for time and cross dependence assumptions between claim
  counts in ratemaking models.
\newblock {\em Insurance: Mathematics and Economics}, 83:161--169.

\bibitem[\protect\astroncite{Boucher}{2023}]{boucher2022}
Boucher, J.~P. (2023).
\newblock Bonus-malus scale models: Creating artificial past claims history.
\newblock {\em Annals of Actuarial Science}, 17(1):36--62.

\bibitem[\protect\astroncite{Boucher and Inoussa}{2014}]{boucher2014}
Boucher, J.~P. and Inoussa, R. (2014).
\newblock A posteriori ratemaking with panel data.
\newblock {\em ASTIN Bulletin: The Journal of the IAA}, 44(3):587--612.

\bibitem[\protect\astroncite{De~Jong et~al.}{2008}]{jong}
De~Jong, P., Heller, G.~Z., et~al. (2008).
\newblock Generalized linear models for insurance data.
\newblock {\em Cambridge Books}.

\bibitem[\protect\astroncite{Delong et~al.}{2021}]{delong2020}
Delong, {\L}., Lindholm, M., and W{\"u}thrich, M.~V. (2021).
\newblock Making tweedie’s compound poisson model more accessible.
\newblock {\em European Actuarial Journal}, 11:185--226.

\bibitem[\protect\astroncite{Denuit et~al.}{2021}]{denuit2021autocalibration}
Denuit, M., Charpentier, A., and Trufin, J. (2021).
\newblock Autocalibration and tweedie-dominance for insurance pricing with
  machine learning.
\newblock {\em Insurance: Mathematics and Economics}, 101:485--497.

\bibitem[\protect\astroncite{Denuit et~al.}{2007}]{denuit2007}
Denuit, M., Mar{\'e}chal, X., Pitrebois, S., and Walhin, J.~F. (2007).
\newblock {\em Actuarial modeling of claim counts: Risk classification,
  credibility and bonus-malus systems}.
\newblock John Wiley \& Sons.

\bibitem[\protect\astroncite{Frees et~al.}{2014}]{frees2014}
Frees, E.~W., Derrig, R.~A., and Meyers, G. (2014).
\newblock {\em Predictive modeling applications in actuarial science},
  volume~1.
\newblock Cambridge University Press.

\bibitem[\protect\astroncite{Hastie et~al.}{2009}]{hastie2009elements}
Hastie, T., Tibshirani, R., and Friedman, J.~H. (2009).
\newblock {\em The elements of statistical learning: data mining, inference,
  and prediction}, volume~2.
\newblock Springer.

\bibitem[\protect\astroncite{Hastie et~al.}{2015}]{statilearn}
Hastie, T., Tibshirani, R., and Wainwright, M. (2015).
\newblock {\em Statistical learning with sparsity: the lasso and
  generalizations}.
\newblock CRC press.

\bibitem[\protect\astroncite{Jeong and Valdez}{2020}]{valdez2020}
Jeong, H. and Valdez, E.~A. (2020).
\newblock Predictive compound risk models with dependence.
\newblock {\em Insurance: Mathematics and Economics}, 94:182--195.

\bibitem[\protect\astroncite{Lee and Shi}{2019}]{shi2019}
Lee, G.~Y. and Shi, P. (2019).
\newblock A dependent frequency--severity approach to modeling longitudinal
  insurance claims.
\newblock {\em Insurance: Mathematics and Economics}, 87:115--129.

\bibitem[\protect\astroncite{Lemaire}{1995}]{lemaire}
Lemaire, J. (1995).
\newblock {\em Bonus-malus systems in automobile insurance}, volume~19.
\newblock Springer science \& business media.

\bibitem[\protect\astroncite{Oh et~al.}{2020}]{oh2020bonus}
Oh, R., Shi, P., and Ahn, J.~Y. (2020).
\newblock Bonus-malus premiums under the dependent frequency-severity modeling.
\newblock {\em Scandinavian Actuarial Journal}, 2020(3):172--195.

\bibitem[\protect\astroncite{Pechon et~al.}{2019}]{pechon2019multivariate}
Pechon, F., Denuit, M., and Trufin, J. (2019).
\newblock Multivariate modelling of multiple guarantees in motor insurance of a
  household.
\newblock {\em European Actuarial Journal}, 9:575--602.

\bibitem[\protect\astroncite{Shi and Valdez}{2014}]{shivaldez2014}
Shi, P. and Valdez, E.~A. (2014).
\newblock Longitudinal modeling of insurance claim counts using jitters.
\newblock {\em Scandinavian Actuarial Journal}, 2014(2):159--179.

\bibitem[\protect\astroncite{Shi and Yang}{2018}]{shi2018pair}
Shi, P. and Yang, L. (2018).
\newblock Pair copula constructions for insurance experience rating.
\newblock {\em Journal of the American Statistical Association},
  113(521):122--133.

\bibitem[\protect\astroncite{Turcotte and Boucher}{2023}]{turcotte2022gamlss}
Turcotte, R. and Boucher, J.~P. (2023).
\newblock Gamlss for longitudinal multivariate claim count models.
\newblock {\em North American Actuarial Journal}, pages 1--24.

\bibitem[\protect\astroncite{Verschuren}{2021}]{verschuren2021predictive}
Verschuren, R.~M. (2021).
\newblock Predictive claim scores for dynamic multi-product risk classification
  in insurance.
\newblock {\em ASTIN Bulletin: The Journal of the IAA}, 51(1):1--25.

\end{thebibliography}

\end{document}